\documentclass[longauth]{aa}
\usepackage{graphicx}
\usepackage{txfonts}
\usepackage[colorlinks=true,citecolor=blue]{hyperref}
\makeatletter
\renewcommand*\aa@pageof{, page \thepage{} of \pageref*{LastPage}}
\makeatother

\begin{document}

   \title{Wolf~327b: A new member of the pack of ultra-short-period super-Earths around M dwarfs 
   }
   
   \author{F.~Murgas\inst{\ref{iac},\ref{ull}}
          \and
          E.~Pall\'{e}\inst{\ref{iac},\ref{ull}}
          \and
          J.~Orell-Miquel\inst{\ref{iac},\ref{ull}}
          \and
          I.~Carleo\inst{\ref{iac},\ref{ull}}
          \and
          L.~Pe\~{n}a-Mo\~{n}ino\inst{\ref{iaa}}
          \and
          M.~P\'{e}rez-Torres\inst{\ref{iaa}, \ref{capa}, \ref{euc}}
          \and
          C.\,N.~Watkins\inst{\ref{cfa}}
          \and
          S.~V.~Jeffers\inst{\ref{mpgott}}
          \and
          M.~Azzaro\inst{\ref{caha}}
          \and
          K.~Barkaoui\inst{\ref{uliege}, \ref{mit}, \ref{iac}}
          \and
          A.\,A.~Belinski\inst{\ref{sai}}
          \and
          J.\,A.~Caballero\inst{\ref{cabinta}}
          \and
          D.~Charbonneau\inst{\ref{cfa}}
          \and
          D.\,V.~Cheryasov\inst{\ref{sai}}
          \and
          D.\,R.~Ciardi\inst{\ref{caltech}}
          \and
          K.\,A.~Collins\inst{\ref{cfa}}
          \and
          M.~Cort\'{e}s-Contreras\inst{\ref{ucm}}
          \and
          J.~de~Leon\inst{\ref{utokyo}}
          \and
          C.~Duque-Arribas\inst{\ref{ucm}}
          \and
          G.~Enoc\inst{\ref{iac},\ref{ull}}
          \and
          E.~Esparza-Borges\inst{\ref{iac},\ref{ull}}
          \and
          A.~Fukui\inst{\ref{komaba}, \ref{iac}}
          \and
          S.~Gerald\'ia-Gonz\'alez\inst{\ref{iac},\ref{ull}}
          \and
          E.\,A.~Gilbert\inst{\ref{jetlab}}
          \and
          A.\,P.~Hatzes\inst{\ref{taut}}
          \and
          Y.~Hayashi\inst{\ref{utokyo}}
          \and
          Th.~Henning\inst{\ref{mpiaa}}
          \and
          E.~Herrero\inst{\ref{ieec}}
          \and
          J.\,M.~Jenkins\inst{\ref{nasaames}}
          \and
          J.~Lillo-Box\inst{\ref{cabinta}}
          \and
          N.~Lodieu\inst{\ref{iac},\ref{ull}}
          \and
          M.\,B.~Lund\inst{\ref{caltech}}
          \and
          R.~Luque\inst{\ref{uchicago}}
          \and
          D.~Montes\inst{\ref{ucm}}
          \and
          E.~Nagel\inst{\ref{ugottingen}}
          \and
          N.~Narita\inst{\ref{komaba}, \ref{actokyo}, \ref{iac}}
          \and
          H.~Parviainen\inst{\ref{iac},\ref{ull}}
          \and
          A.\,S.~Polanski\inst{\ref{ukansas}}
          \and
          S.~Reffert\inst{\ref{lsw}}
          \and
          M.~Schlecker\inst{\ref{uarizona}}
          \and
          P.~Sch{\"o}fer\inst{\ref{iaa}}
          \and
          R.\,P.~Schwarz\inst{\ref{cfa}}
          \and
          A.~Schweitzer\inst{\ref{uham}}
          \and
          S.~Seager\inst{\ref{kavli}, \ref{mit}, \ref{mitaero}}
          \and
          K.\,G.~Stassun\inst{\ref{uvander}}
          \and
          H.\,M.~Tabernero\inst{\ref{ucm}}
          \and
          Y.~Terada\inst{\ref{iaa_taiwan}, \ref{ntu}}
          \and
          J.\,D.~Twicken\inst{\ref{seti}, \ref{nasaames}}
          \and
          S.~Vanaverbeke\inst{\ref{vereniging},\ref{leuvenmat},\ref{astrolab}}
          \and
          J.\,N.~Winn\inst{\ref{princeton}}
          \and
          R.~Zambelli\inst{\ref{sal}}
          \and
          P.\,J.~Amado\inst{\ref{iaa}}
          \and
          A.~Quirrenbach\inst{\ref{lsw}}
          \and
          A.~Reiners\inst{\ref{ugottingen}}
          \and
          I.~Ribas\inst{\ref{ice},\ref{ieec}}
          }

   \institute{
Instituto de Astrof\'isica de Canarias (IAC), calle V\'ia L\'actea s/n, 38205 La Laguna, Tenerife, Spain \label{iac} \\
              \email{fmurgas@iac.es} 
\and
Departamento de Astrof\'isica, Universidad de La Laguna (ULL), 38206 La Laguna, Tenerife, Spain \label{ull}
\and
Instituto de Astrof\'{i}sica de Andaluc\'{i}a (IAA-CSIC), Glorieta de la Astronom\'{i}a s/n, 18008 Granada, Spain\label{iaa}
\and
Center for Astroparticles and High Energy Physics (CAPA), Universidad de Zaragoza, E-50009 Zaragoza, Spain \label{capa}
\and
School of Sciences, European University Cyprus, Diogenes street, Engomi, 1516 Nicosia, Cyprus \label{euc}
\and
Center for Astrophysics \textbar ~ Harvard \& Smithsonian, 60 Garden Street, Cambridge, MA 02138, USA\label{cfa}
\and
Max Planck Institute for Solar System Research, Justus-von-Liebig-Weg 3, D-37077, G\"ottingen, Germany\label{mpgott}
\and
Centro Astron\'onomico Hispano en Andaluc\'ia, Observatorio de Calar Alto, Sierra de los Filabres, 04550 G\'ergal, Almer\'ia, Spain\label{caha}
\and
Astrobiology Research Unit, Universit\'e de Li\`ege, 19C All\'ee du 6 Ao\^ut, 4000 Li\`ege, Belgium\label{uliege}
\and
Department of Earth, Atmospheric and Planetary Science, Massachusetts Institute of Technology, 77 Massachusetts Avenue, Cambridge, MA 02139, USA\label{mit}
\and
Sternberg Astronomical Institute Lomonosov Moscow State University 119992, Universitetskii prospekt, 13, Moscow, Russia\label{sai}
\and
Centro de Astrobiolog\'{i}a (CSIC-INTA), ESAC campus, 28692 Villanueva de la Ca\~{n}ada, Madrid, Spain \label{cabinta}
\and
NASA Exoplanet Science Institute, IPAC, California Institute of Technology, Pasadena, CA 91125 USA\label{caltech}
\and
Departamento de F\'isica de la Tierra y Astrof\'isica and IPARCOS-UCM (Instituto de F\'isica de Part\'iculas y del Cosmos de la UCM), Facultad de Ciencias F\'isicas, Universidad Complutense de Madrid, 28040, Madrid, Spain\label{ucm}
\and
Department of Multi-Disciplinary Sciences, Graduate School of Arts and Sciences, The University of Tokyo, 3-8-1 Komaba, Meguro, Tokyo 153-8902, Japan\label{utokyo}
\and
Komaba Institute for Science, The University of Tokyo, 3-8-1 Komaba, Meguro, Tokyo 153-8902, Japan\label{komaba}
\and
Jet Propulsion Laboratory, California Institute of Technology, 4800 Oak Grove Drive, Pasadena, CA 91109, USA\label{jetlab}
\and
Th\"uringer Landessternwarte Tautenburg, Sternwarte 5, D-07778 Tautenburg, Germany\label{taut}
\and
Max-Planck-Institut f\"ur Astronomie, K\"onigstuhl 17, D-69117 Heidelberg, Germany\label{mpiaa}
\and
Institut d’Estudis Espacials de Catalunya (IEEC), c/ Gran Capit\`a 2-4, 08034 Barcelona, Spain\label{ieec}
\and
NASA Ames Research Center, Moffett Field, CA 94035, USA\label{nasaames}
\and
Department of Astronomy and Astrophysics, University of Chicago, Chicago, IL 60637, USA\label{uchicago}
\and
Institut f\"ur Astrophysik und Geophysik, Georg-August-Universit\"at, Friedrich-Hund-Platz 1, 37077 G\"ottingen, Germany\label{ugottingen}
\and
Astrobiology Center, 2-21-1 Osawa, Mitaka, Tokyo 181-8588, Japan\label{actokyo}
\and
Department of Physics and Astronomy, University of Kansas, Lawrence, KS 66045, USA\label{ukansas}
\and
Landessternwarte, Zentrum f\"ur Astronomie der Universit\"at Heidelberg, K\"onigstuhl 12, 69117 Heidelberg, Germany\label{lsw}
\and
Steward Observatory and Department of Astronomy, The University of Arizona, Tucson, AZ 85721, USA\label{uarizona}
\and
Hamburger Sternwarte, Universit\"at Hamburg, Gojenbergsweg 112, 21029 Hamburg, Germany\label{uham}
\and
Department of Physics and Kavli Institute for Astrophysics and Space Research, Massachusetts Institute of Technology, Cambridge, MA 02139, USA\label{kavli}
\and
Department of Aeronautics and Astronautics, MIT, 77 Massachusetts Avenue, Cambridge, MA 02139, USA\label{mitaero}
\and
Department of Physics and Astronomy, Vanderbilt University, Nashville, TN 37235, USA\label{uvander}
\and
Institute of Astronomy and Astrophysics, Academia Sinica, P.O. Box 23-141, Taipei 10617, Taiwan, R.O.C.\label{iaa_taiwan}
\and
Department of Astrophysics, National Taiwan University, Taipei 10617, Taiwan, R.O.C.\label{ntu}
\and
SETI Institute, Mountain View, CA 94043 USA\label{seti}
\and
Vereniging Voor Sterrenkunde (VVS), Oostmeers 122 C, 8000 Brugge, Belgium\label{vereniging}
\and
Centre for Mathematical Plasma-Astrophysics, Department of Mathematics, KU Leuven, Celestijnenlaan 200B, 3001 Heverlee, Belgium\label{leuvenmat}
\and
Public Observatory ASTROLAB IRIS, Provinciaal Domein “De Palingbeek”, Verbrandemolenstraat 5, 8902 Zillebeke, Ieper, Belgium\label{astrolab}
\and
Department of Astrophysical Sciences, Princeton University, 4 Ivy Lane, Princeton, NJ 08544, USA\label{princeton}
\and
Societ\`a Astronomica Lunae, Castelnuovo Magra, Italy\label{sal}
\and
Institut de Ci\`encies de l'Espai (ICE, CSIC), Campus UAB, c/ de Can Magrans s/n, 08193 Bellaterra, Barcelona, Spain\label{ice}
}
\date{Received 1 December 2023 / Accepted 21 January 2024}

 
  \abstract{Planets with orbital periods shorter than 1 day are rare and have formation histories that are not completely understood. Small ($R_\mathrm{p} < 2\; R_\oplus$) ultra-short-period (USP) planets are highly irradiated, probably have rocky compositions with high bulk densities, and are often found in multi-planet systems. Additionally, USP planets found around small stars are excellent candidates for characterization using present-day instrumentation. Of the current full sample of approximately 5500 confirmed exoplanets, only 130 are USP planets and around 40 have mass and radius measurements. Wolf~327 (TOI-5747) is an M dwarf ($R_\star = 0.406 \pm 0.015 \; R_\odot$, $M_\star = 0.405 \pm 0.019 \; M_\odot$, $T_{\mathrm{eff}}=3542 \pm 70$\,K, and $V = 13$\,mag) located at a distance $d = 28.5$\,pc. NASA's planet hunter satellite, TESS, detected transits in this star with a period of 0.573\,d (13.7\,h) and with a transit depth of 818\,ppm. Ground-based follow-up photometry, high resolution imaging, and radial velocity (RV) measurements taken with the CARMENES spectrograph confirm the presence of this new USP planet. Wolf~327b is a super-Earth with a radius of $R_\mathrm{p} = 1.24 \pm 0.06 \; R_\oplus$ and a mass of $M_\mathrm{p} = 2.53 \pm 0.46 \; M_\oplus$, yielding a bulk density of $7.24 \pm 1.66 $\,g\,cm$^{-3}$ and thus suggesting a rocky composition. Owing to its close proximity to its host star ($a = 0.01$\,au), Wolf~327b has an equilibrium temperature of $996 \pm 22$\,K. This planet has a mass and radius similar to K2-229b, a planet with an inferred Mercury-like internal composition. Planet interior models suggest that Wolf~327b has a large iron core, a small rocky mantle, and a negligible (if any) H/He atmosphere.}


   \maketitle
%

\section{Introduction}
\label{Sec:Intro}
One of the most uncommon types of exoplanets discovered to date are ultra-short-period (USP) planets. USP planets are defined arbitrarily as planets with orbital periods shorter than one day \citep{Sahu2006,SanchisOjeda2013}. Owing to their short planet-to-star distance, they experience levels of irradiation on the order of hundreds or thousands of times stronger than Earth's, and are thus subject to extreme temperatures. These planets typically have small sizes, with planetary radii up to $R_\mathrm{p} < 2 \; R_\oplus$ \citep{Winn2018}, and appear to have mostly Earth-like compositions \citep{Dai2019}, some of them even exhibiting iron-enhanced densities \citep[e.g.,][]{Rappaport2013,Santerne2018,PriceRogers2020}. USP planets have certain characteristics that put them in a distinct category of exoplanets. For example, unlike hot Jupiters, the presence of USP planets does not strongly correlate with the metallicity of their parent star \citep{Winn2017}. Another characteristic that separates USP planets from hot Jupiters is that the former are often found in multi-planetary systems \citep{SanchisOjeda2014, Winn2018}. Furthermore, the abundance of USP planets seems to depend on the spectral type of their host star; \cite{SanchisOjeda2014} computed the USP planet occurrence rates for different stellar spectral types and found $0.15 \pm 0.05$\% for F-type stars, $0.51 \pm 0.07$\% for G-type stars, and $1.1 \pm 0.4$\% for M-type stars. Although the measured occurrence of USP planets is higher for M dwarfs than for F dwarfs, that difference is of modest statistical significance (about 2$\sigma$). More significant is the higher occurrence around G dwarfs compared to F dwarfs (about 4$\sigma$), making it seem reasonable that the occurrence decreases as a function of stellar mass, in general.

The formation pathway of USP planets is still unknown. Because of the high irradiation levels they experience, USP planets typically lie within the region where dust is sublimated, which means that they probably had to form farther away from their current orbit. Proposed formation mechanisms include photo-evaporation of sub-Neptune planets \citep[e.g.,][]{Lundkvist2016, LeeChiang2017} and migration of rocky planets \citep[e.g.,][]{Petrovich2019}.

Highly irradiated rocky USP planets may offer a unique opportunity to study the surface and internal composition of exoplanets with current instrumentation. Planets with surface temperatures above $\sim$1100\,K will probably have large molten regions since this temperature range marks the melting point of rocks for Earth-like compositions \citep{Katz2003}. This surface temperature range can be achieved by USP planets owing to their short orbital periods and because they are most probably tidally locked to their star. Although USP rocky planets probably lose their primordial atmospheres in the early stages of their formation owing to the effects of the stellar activity and winds of their host star, some volatile elements associated with the magma oceans located in the planet's dayside may be detected in emission \citep{Ito2015,Nguyen2020,Nguyen2022}. 

Currently, no confirmed detection of a USP rocky planet atmosphere has been reported; however, the canonical USP rocky exoplanet 55~Cnc~e \citep[$P=0.73$ d, $R_\mathrm{p} = 1.87 \; R_\oplus$, and $M_\mathrm{p} = 7.99 \; M_\oplus$;][]{McArthur2004,Dawson2010,Winn2011,Demory2011} is a promising candidate for the detection of a secondary atmosphere with an origin associated with its magma ocean. The depths of the secondary eclipses of 55~Cnc~e have shown evidence of variability at both optical \citep{MeierValdes2022,Demory2023,MeierValdes2023} and infrared wavelengths \citep{Demory2016,Tamburo2018}, while the primary transit depths do not exhibit any significant variability \citep{Tamburo2018,MeierValdes2023}. \cite{Heng2023} proposed an explanation for 55~Cnc~e observations in which a secondary atmosphere is produced by geochemical outgassing events on the dayside of the planet; this gas envelope is transient owing to atmospheric escape processes. In this scenario an eclipse with a depth consistent with zero is produced when there is no secondary atmosphere; when the outgassed atmosphere accumulates in the dayside, this then has the effect of increasing the eclipse depth at optical wavelengths owing to a combination of Rayleigh scattering and thermal emission; and finally fluctuations in temperature are responsible for the variable infrared eclipse depth. Future JWST observations may detect evidence for this proposed scenario for 55~Cnc~e; however, more observations are needed to explore whether other rocky USP planets also have variable infrared secondary eclipse depths, such observations perhaps opening the door to the study of secondary atmospheres in exoplanets.

To further improve our understanding of the formation and evolution of USP planets, it is important to increase the sample of this type of object with accurately measured radii and masses. In this paper we report the detection of a USP planet around the star Wolf~327. This star is an M2.5-type star located at $d = 28.5$\,pc. The transit events were discovered by the Transiting Exoplanet Survey Satellite (TESS) as part of its search for transiting planets. This paper is organized as follows: in Sect.\ \ref{Sec:SpaceObs} and \ref{Sec:GroundObs} we describe TESS and the ground-based data of this target; Sect. \ref{Sec:StarProper} presents the stellar properties of the host star; Sect. \ref{Sec:AnalysisResults} describes our data analysis methods and results; and Sect. \ref{Sec:Discussion} discusses our results and puts Wolf~327b in the context of known USP planets; and finally in Sect. \ref{Sec:Conclusions} we present our conclusions.

\section{TESS observations}
\label{Sec:SpaceObs}
The transiting planet candidate was discovered by TESS \citep{Ricker2014}. TESS takes photometric data continuously for $\sim$27 d and sends data to Earth every $\sim$13.7 d. Wolf~327 (TIC~4918918, TOI-5747) was observed at 2-minute cadence in sector 21 from 2020 January 21 to 2020 February 18 and in sector 48 from 2022 January 28 to 2022 February 26. Once the data are received, the TESS Science Processing Operations Center \citep[SPOC,][]{Jenkins2016} at the NASA Ames Research Center generates simple aperture-photometry \citep[SAP,][]{Morris2020} light curves that are then processed to remove systematic effects using the Presearch Data Conditioning (PDC) pipeline module \citep{Smith2012,Stumpe2012,Stumpe2014}. The light curves are then scanned for transit-like signals with a wavelet-based adaptive noise-compensating matched filter \citep{Jenkins2002,Jenkins2010,JenkinsJM2020}. A limb-darkened transiting planet model is fitted \citep{Li2019}, and a suite of false-positive detection tests are applied to rule out certain non-planetary scenarios \citep{Twicken2018}. Once this transit search and model-fitting process is completed, the candidate reports are vetted by the TESS Science Office (TSO) at Massachusetts Institute of Technology (MIT), and the planet candidates are announced to the community and assigned a TESS object of interest (TOI) number. On 2022 September 1, TSO announced the detection of a transiting planet candidate around Wolf~327. A transit signature with an orbital period of 0.573\,d ($\sim$13.7\,h) was detected in the SPOC search of the combined light curve from the two sectors on 2022 May 26. The transit signature passed all the diagnostic tests presented in the SPOC Data Validation reports. The SPOC Data Validation process performs difference imaging (out-of-transit baseline image minus in-transit image) and centroid offset analysis as described by \cite{Twicken2018} to establish the location of the transit source with respect to the location of the target star. A mean difference image is produced in each sector (with transits) for a given threshold crossing event and then a centroid based on the pixel response function is computed to determine the location of the transit source. Afterwards a centroid offset is determined by subtracting the location of the target star, corrected for proper motion. Wolf~327 was observed in two sectors (sectors 21 and 48); for targets observed in multiple sectors SPOC computes a mean centroid offset by robustly averaging over the centroid offsets associated with the individual sectors. The mean offset located the source of the transit at $5.13 \pm 4.20 \arcsec$ from the coordinates of Wolf~327. There are no other TESS Input Catalog (TIC) objects within $3\sigma$ of the mean centroid offset, hence all TIC objects are excluded as potential sources of the transit signal. After passing all the diagnostics tests performed by SPOC, the candidate was assigned TOI number TOI-5747.01 by the TESS Science Office.

\begin{figure*}
   \centering
   \includegraphics[width=\hsize]{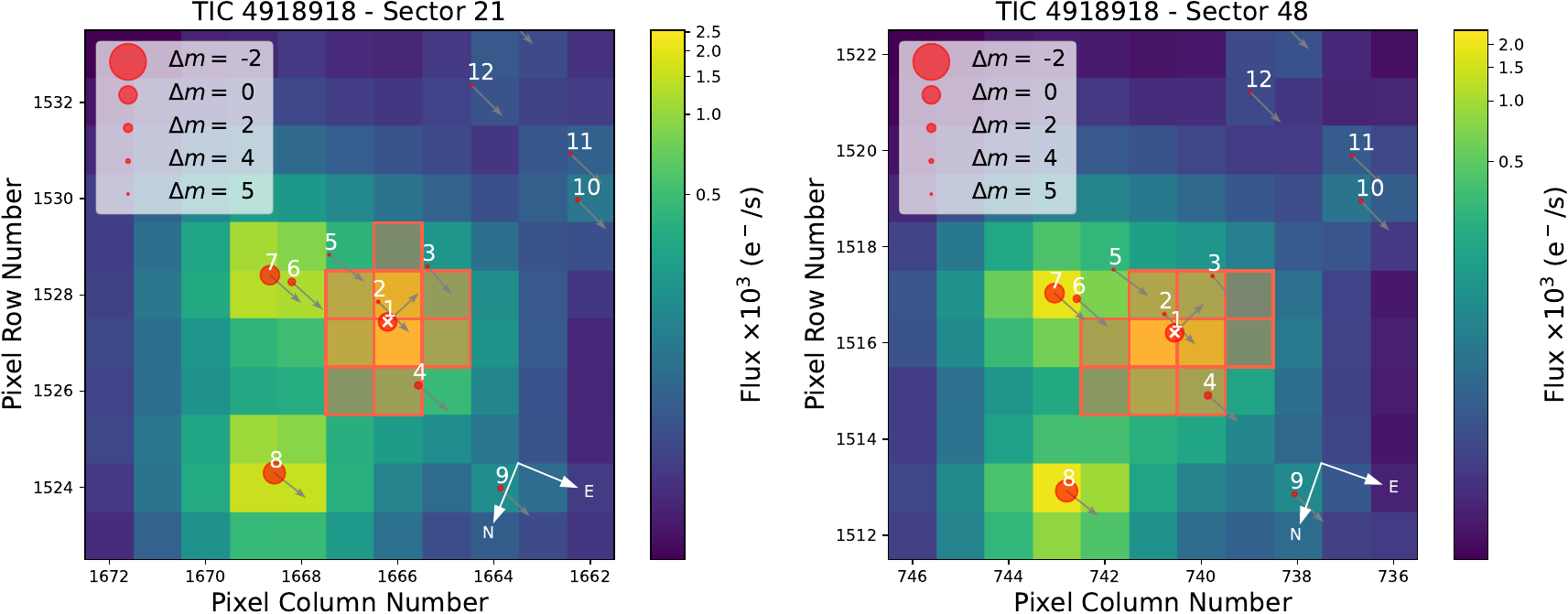}
   \caption{\texttt{Tpfplotter} \citep{Aller2020} target pixel file (TPF) images of Wolf~327 for sectors 21 and 48. The red squares correspond to the TESS aperture used to compute the photometry, the size of the red circles represent the \textit{Gaia} DR3 \citep{GaiaDR3} magnitudes of the stars, and the gray arrows shows the \textit{Gaia} DR3 proper motion directions of each star inside the field of view. Wolf~327b is positioned at the center (star 1), star 7 (TIC~4918919) is a known eclipsing binary.}
    \label{Fig:TESS_TPFs}
\end{figure*}

In this study we have analyzed TESS PDCSAP photometry of Wolf~327 from sectors 21 and 48, both taken with a 2-minute cadence. Figure \ref{Fig:TESS_TPFs} shows the TESS target pixel file (TPF) produced with \texttt{tpfplotter},\footnote{\url{https://github.com/jlillo/tpfplotter}} the figure shows the field around the target, highlighting in red the apertures used to compute the light curves. TESS apertures were computed by SPOC's compute optimal aperture, which selects the photometric aperture pixels for each target and sector. The code also calculates a ``crowding metric'' for each photometric aperture; this is the fraction of light in the photometric aperture due to the target star after background subtraction. In sector 21 the crowding metric was 0.9259, therefore 7.41\% of the median flux level was subtracted on each cadence to account for light from other nearby stars in the PDCSAP light curve. In sector 48 the crowding metric was 0.9060, hence 9.40\% of the median flux level was subtracted to account for dilution. Otherwise, the transit depths would be artificially small. This dilution correction is based on a scalar average of each sector and is not dynamic.

\begin{figure*}
   \centering
   \includegraphics[width=\hsize]{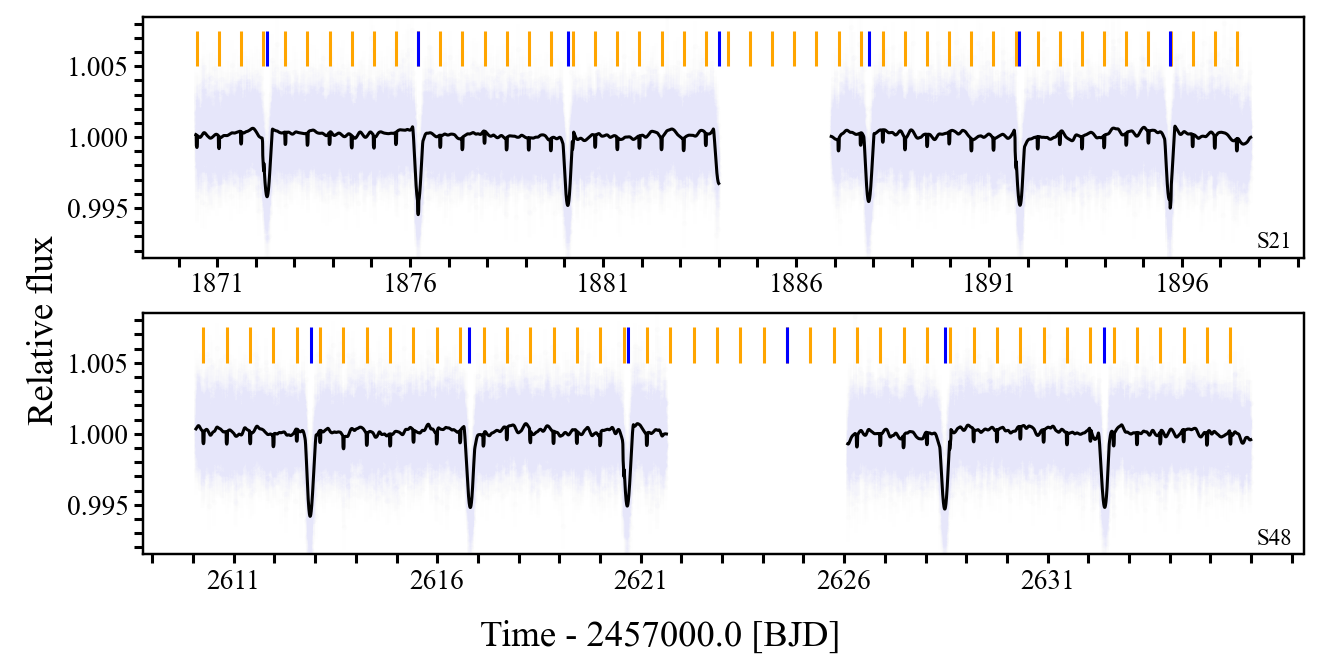}
   \caption{TESS PDCSAP photometry for sectors 21 and 48. The deep transit events occurring with a period of $\sim$3.89 d are from the known eclipsing binary TIC~4918919. The solid black line shows the best-fitting model from the joint data fit described in Sect. \ref{Sec:JointFit}. The vertical lines represent the central time of the transit of the USP Wolf~327b (orange) and the eclipsing times of the EB TIC~4918919 (blue).}
    \label{Fig:Wolf327b_TESS_LC}
\end{figure*}

Figure \ref{Fig:Wolf327b_TESS_LC} shows the TESS light curves for sectors 21 and 48; both time series clearly show the eclipses of the nearby star TIC~4918919 ($V=12.1$ mag, $J=10.7$ mag). TIC~4918919 is located at a projected separation of $r=55 \arcsec$ from Wolf~327. This nearby star is a known eclipsing binary classified by \textit{Gaia} (\textit{Gaia} DR3 796185373590623360), and TESS observes an eclipse every $\sim$3.89 d. This eclipsing binary is a background star (\textit{Gaia} DR3 parallax $\pi = 1.53 \pm 0.02$ mas) and is not bound to Wolf~327 (\textit{Gaia} DR3 parallax $\pi = 35.00 \pm 0.03$ mas). Although this star is positioned outside TESS's photometric aperture for both sectors (see Fig.\ \ref{Fig:TESS_TPFs}) owing to its brightness, part of its point spread function (PSF) is included inside the aperture used to integrate the flux of Wolf~327. As previously mentioned, the constant flux contamination level of TIC~4918919 is taken into account by the PDCSAP pipeline; however, this dilution correction is not dynamic and the eclipses are still seen in the final TESS light curve. In Sect. \ref{Sec:JointFit} we describe the steps taken to account for the contributions of the transiting candidate and eclipsing binary present in both TESS time series.

\section{Ground-based observations}
\label{Sec:GroundObs}

\subsection{Seeing-limited photometry}
The TESS pixel scale is $\sim 21\arcsec$ pixel$^{-1}$ and photometric apertures typically extend out to roughly $1 \arcmin$, generally causing multiple stars to blend in the TESS photometric aperture. To determine the true source of the TESS detection, we acquired ground-based time-series follow-up photometry of the field around Wolf~327 as part of the validation process of the candidate.

\subsubsection{Long-term photometric monitoring}
We searched for long-term photometric data of Wolf~327 in archival databases. We found a time series taken by the All-Sky Automated Survey for Supernovae \citep[ASAS-SN,][]{Shappee2014}. ASAS-SN is a survey dedicated to finding bright supernovae and uses an array of 24 robotic telescopes located in several countries (USA, Chile, South Africa, China). The project provides light curves for stars with magnitudes between 9 and 18 in $g$ and $V$ filters through their Sky Patrol portal.\footnote{\url{https://asas-sn.osu.edu/photometry}} We searched the ASAS-SN portal for Wolf~327 photometry and found $\sim$200 epochs taken with the $V$ filter, spanning a time baseline of $\sim$2060 d. The median values of the photometric measurements are $\bar{V}=12.98$ mag with a median photometric uncertainty of $\sigma_V = 0.02$ mag; the standard deviation of the time series is $\sigma_{\mathrm{dev}} = 0.016$ mag. During the period of time in which the star was visible, the cadence of the observations was roughly one visit every 3 d. 

A long term photometric campaign of Wolf~327 was carried out from April to December 2023 with the 0.8 m Joan Or\'{o} telescope \citep[TJO,][]{Colome2010} at the Montsec Observatory in Lleida, Spain. We obtained a total of 353 images on 61 different nights with an exposure time of 60\,s each using the Johnson $R$ filter of the LAIA imager, a 4k$\times$4k with a field of view of 30$\arcmin$ and a scale of $0\farcs4$ pixel$^{-1}$. The images were calibrated with darks, bias, and flat fields with the ICAT pipeline \citep{Colome2006} of the TJO. The differential photometry was extracted with \texttt{AstroImageJ} \citep{Collins:2017} using the aperture size that minimized the root-mean-square (rms) of the resulting relative fluxes, and a selection of the ten brightest comparison stars in the field that did not show variability. Then, we used our own pipelines to remove outliers and measurements affected by poor observing conditions or with a low signal-to-noise ratio (S/N). The resulting rms of the differential photometry from the TJO in the $R$ filter is $\sim$11 ppt.

Owing to the relatively large photometric uncertainties and the cadence of the observations, it is difficult to detect the transit events in the ASAS-SN and TJO data; however, the long-term monitoring of Wolf~327 provides insights into the level of stellar activity of the star and helps to establish the rotation period of Wolf~327 (see Sect.\ \ref{Sec:StellarRotation}).

\subsubsection{MuSCAT2 photometry}

Wolf~327 was observed on the night of 2022 November 6 with the 1.5\,m Telescopio Carlos S\'{a}nchez (TCS) at Teide Observatory, Spain. The data were acquired by the TCS MuSCAT2 imager \citep{Narita2019}. This instrument is equipped with four cameras and is capable of obtaining simultaneous images in $g'$, $r'$, $i'$, and $z$-short with little overhead time. 

During the observations the exposure times were set to 10\,s for the four bands. Standard data reduction and calibrations were performed by the MuSCAT2 pipeline \citep{Parviainen2019}; in addition to basic reduction, this pipeline also optimizes the photometric aperture while fitting a transit model (including systematic effects) to the time series. 

Since the data covered an egress of the transit and not a full event, they were not used in the final fit. However, the MuSCAT2 observations were useful to detect hints of an egress on Wolf~327, confirming the transit event on the target star and ruling out the presence of hitherto unrecognized eclipsing binaries located inside TESS's photometric aperture that could mimic a signal similar to that caused by a planetary transit.

\subsubsection{LCOGT photometry}
Wolf~327 was included as part of the TESS Follow-up Observing Program \citep[TFOP;][]{Collins:2019}\footnote{\url{https://tess.mit.edu/followup}}. The on-target follow-up light curves are also used to place constraints on the transit depth and the TESS ephemeris. To schedule our transit observations we used the {\tt TESS Transit Finder}, which is a customized version of the {\tt Tapir} software package \citep{Jensen:2013}.

We observed three full transit windows of Wolf~327b on 2022 November 19, 2022 December 2 and 2022 December 12 in the Pan-STARRS $z$-short band and two in the Sloan $i'$ band, respectively, using the Las Cumbres Observatory Global Telescope \citep[LCOGT;][]{Brown:2013} 1.0\,m network nodes at Teide Observatory on the island of Tenerife and at McDonald Observatory near Fort Davis, Texas, United States. The images were calibrated with the standard LCOGT {\tt BANZAI} pipeline \citep{McCully:2018} and differential photometric data were extracted using {\tt AstroImageJ}. We used circular photometric apertures with a radius of $4\farcs7$ for the $z$-short band observations and $12\farcs4$ and $5\farcs8$ for the two $i'$ band observations, respectively, to extract the differential photometry. The smaller target star apertures excluded all of the flux from the nearest known neighbor in the \textit{Gaia} DR3 catalog (\textit{Gaia} DR3 796185407950360320), which is $\sim9\arcsec$ southwest of Wolf~327. The observational mode on the night of 2022 December 2 used greater defocusing and thus resulted in a larger photometric aperture that includes flux from the nearest known neighbor. However the neighbor is more than a hundred times fainter than Wolf~327 and insignificantly affects the transit depth. The three light curves are included in the global modeling described in Sect. \ref{Sec:JointFit}.

Two full transit windows and two ingress windows of Wolf~327b were observed simultaneously in Sloan $g'$, $r'$, $i'$, and Pan-STARRS $z$-short bands on 2023 January 19, 2023 January 23, 2023 March 25, and 2023 April 25 using the LCOGT 2\,m Faulkes Telescope North at Haleakala Observatory on Maui, Hawai'i. The telescope is equipped with the MuSCAT3 multi-band imager \citep{Narita:2020}. The images were calibrated using the standard LCOGT {\tt BANZAI} pipeline \citep{McCully:2018}, and photometric data were extracted using {\tt AstroImageJ} \citep{Collins:2017}. We used circular apertures with a radius of $\sim 5\farcs0$ to extract the differential photometry. The two ingress window observations on 2023 January 23 and 2025 March 25 were not included in the global modeling. 

\subsection{High resolution imaging}
As part of our standard process for validating transiting exoplanets to assess the possible contamination by bound or unbound companions on the derived planetary radii \citep{Ciardi2015}, we observed Wolf~327 with high-resolution near-infrared adaptive optics (AO) imaging at Keck Observatory and with optical speckle observations at Sternberg Astronomical Institute.

\subsubsection{Near-infrared AO at Keck}
Observations of Wolf~327 were taken with the NIRC2 instrument on Keck-II behind the natural guide star AO system \citep{Wizinowich2000} on 2023 April 26 in the standard 3-point dither pattern that is used with NIRC2 to avoid the left lower quadrant of the detector, which is typically noisier than the other three quadrants. The dither pattern step size was $3\arcsec$ and was repeated twice, with each dither offset from the previous dither by $0\farcs5$. NIRC2 was used in the narrow-angle mode with a full field of view of $\sim 10\arcsec$ and a pixel scale of approximately $0\farcs0099442$ per pixel. The Keck observations were made in the $Kcont$ filter $(\lambda_o = 2.2706; \Delta\lambda = 0.0296~\mu$m) with an integration time in each filter of 20 s for a total of 180 s. Flat fields were generated from a median average of dark-subtracted dome flats. Sky frames were generated from the median average of the nine dithered science frames; each science image was then sky-subtracted and flat-fielded. The reduced science frames were combined into a single combined image using an intra-pixel interpolation that conserves flux and shifts the individual dithered frames by the appropriate fractional pixels; the final resolution of the combined dithers was determined from the full-width half-maximum (FWHM) of the point spread function ($0\farcs100$). To within the limits of the AO observations, no stellar companions were detected. The final $5\sigma$ limit at each separation was determined from the average of all of the determined limits at that separation and the uncertainty on the limit was set by the root-mean-square dispersion of the azimuthal slices at a given radial distance (Figure \ref{Fig:Wolf327_HighResImaging}, left panel).

\subsubsection{Optical speckle at SAI}
\begin{figure*}
    \centering
    \includegraphics[width=\hsize]{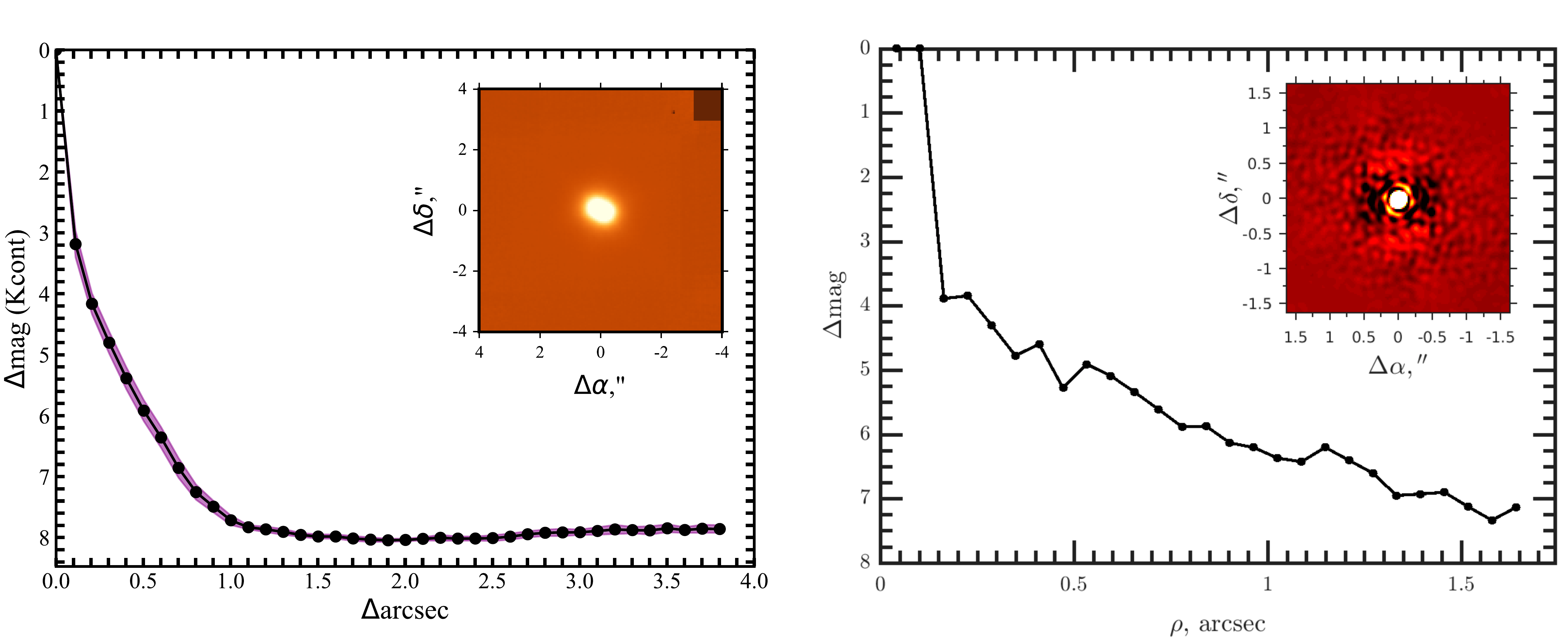}
    \caption{Adaptive optics and speckle imaging of Wolf~327. \textit{Left panel:} Keck NIR AO imaging and sensitivity curve for Wolf~327 taken with the $Kcont$ filter. The black points represent the 5$\sigma$ limits and are separated in steps of 1 FWHM; the purple area represents the 1$\sigma$ azimuthal dispersion of the contrast determinations (see text). The inset image of the primary target shows no additional close-in companions. \textit{Right panel:} $I_\mathrm{c}$ band speckle sensitivity curve of Wolf~327 taken with the SPeckle Polarimeter (SPP) at the Caucasian Observatory. No stellar companions are detected with magnitudes brighter than $\Delta I_\mathrm{c}=3.9$ and 6.2 at $\rho = 0\farcs25$ and $1\farcs0$ respectively. The inset shows the speckle autocorrelation function.}
    \label{Fig:Wolf327_HighResImaging}
\end{figure*}

Wolf~327 was observed on 2023 January 17 with the speckle polarimeter on the 2.5 m telescope at the Caucasian Observatory of Sternberg Astronomical Institute (SAI) of Lomonosov Moscow State University. The speckle polarimeter uses a high-speed low-noise CMOS detector Hamamatsu ORCA--quest \citep{Strakhov2023}. The atmospheric dispersion compensator was active, which allowed use of the $I_\mathrm{c}$ band. The respective angular resolution is $0\farcs083$, while the long-exposure atmospheric seeing was $0\farcs8$. We did not detect any stellar companions brighter than $\Delta I_\mathrm{c}=3.9$ and 6.2 at $\rho = 0\farcs25$ and $1\farcs0$, respectively, where $\rho$ is the separation between the source and the potential companion. The sensitivity curve obtained is shown in Fig.\ \ref{Fig:Wolf327_HighResImaging} (right panel).

\subsubsection{Gaia assessment}
In addition to employing high-resolution imaging, we have utilized \textit{Gaia} DR3 to identify possible wide companions bound to Wolf~327. Typically, these stars are already cataloged in the TESS Input Catalog, and their impact on the transit events has been factored into the analysis of the transits and the associated derived parameters. From a comparison of parallaxes and proper motions \citep[e.g.,][]{mugrauer2020,mugrauer2021,mugrauer2022,mugrauer2023}, our investigation reveals no additional widely separated companions identified by \textit{Gaia}.

Furthermore, the astrometry from \textit{Gaia} offers supplementary insights into the potential existence of close companions that might have evaded detection by both \textit{Gaia} and high-resolution imaging. The \textit{Gaia} Renormalized Unit Weight Error (RUWE) serves as a metric, akin to a reduced chi-square, with values approximately $\lesssim 1.4$ indicating that the astrometric solution is compatible with a solitary star. Conversely, RUWE values $\gtrsim 1.4$ suggest an excess of astrometric noise, potentially attributed to the presence of an unseen companion \citep[e.g., ][]{ziegler2020}. Wolf~327, with a RUWE of $1.3$, aligns with this single-star model, suggesting no unseen companions causing excess astrometric noise.

\subsection{CARMENES radial velocity measurements}
Wolf~327 was observed with the CARMENES spectrograph as part of an RV follow-up program of TESS candidates (22B-3.5-006; PI: E.~Pall\'{e}). CARMENES is installed at the 3.5\,m telescope at the Calar Alto Observatory in Almer\'{i}a, Spain. The instrument has two channels: the visible one (VIS, spectral range 0.52--0.96\,$\mu$m) and the near-infrared one (NIR, spectral range 0.96--1.71\,$\mu$m); with an average spectral resolution for both channels of $R = 94\,600$ and $R = 80\,400$, respectively \citep{Quirrenbach2014,Quirrenbach2018}.

We collected a total of 22 CARMENES spectra from 2023 January 6 until 2023 February 6, covering a time baseline of 31 d. The exposure time used to acquire the spectra was 1800\,s. The data reduction was done with the \texttt{CARACAL} pipeline \citep{Caballero2016}, which performs basic data reduction (bias, flat, and cosmic ray corrections) and extracts the spectra using the {\tt FOX} optimal extraction algorithm \citep{Zechmeister2014}. The wavelength calibration is done following the algorithms described in \citet{Bauer2015}. The radial velocity measurements were computed with \texttt{SERVAL}\footnote{\url{https://github.com/mzechmeister/serval}} \citep{Zechmeister2018}. \texttt{SERVAL} produces a template spectrum by co-adding and shifting the observed spectra and computes the RV shift relative to this template using a $\chi^2$ minimization with the RV shift as a free parameter. The RV measurements were corrected for other effects such as barycentric motion, secular acceleration, instrumental drifts, and nightly zero-points \citep[e.g.,][]{Trifonov2018,TalOr2018}.

\texttt{SERVAL} also computes several activity indices, such as spectral line indices (the Na~{\sc i} doublet $\lambda\lambda$589.0\,nm, 589.6\,nm; H$\alpha$ $\lambda$656.2\,nm, and the Ca~{\sc ii} infrared triplet $\lambda\lambda\lambda$849.8\,nm, 854.2\,nm, 866.2\,nm), and other indicators such as the RV chromatic index (CRX) and the differential line width (dLW). These indicators are useful for monitoring the stellar activity of the star and its effect on the RV measurements \citep[cf.][]{Zechmeister2018}.

The uncertainties of the RV measurements derived by \texttt{SERVAL} have a median value of $1.88$\,m\,s$^{-1}$ and a standard deviation of $\sigma_{\rm dev} = 1.43$\,m\,s$^{-1}$ for the VIS channel. In the NIR channel, the median RV uncertainty value is larger at $9.69$\,m\,s$^{-1}$, with a standard deviation of $\sigma_{\rm dev}=4.95$\,m\,s$^{-1}$. The median overall signal-to-noise ratio for the spectra taken with the VIS channel was 329, with a standard deviation of 99 (minimum S/N = 88, maximum S/N = 469). For the NIR spectra, the median S/N was 446, with a standard deviation of 138 (min. S/N = 111, max. S/N = 602).

\begin{figure}
    \centering
    \includegraphics[width=\hsize]{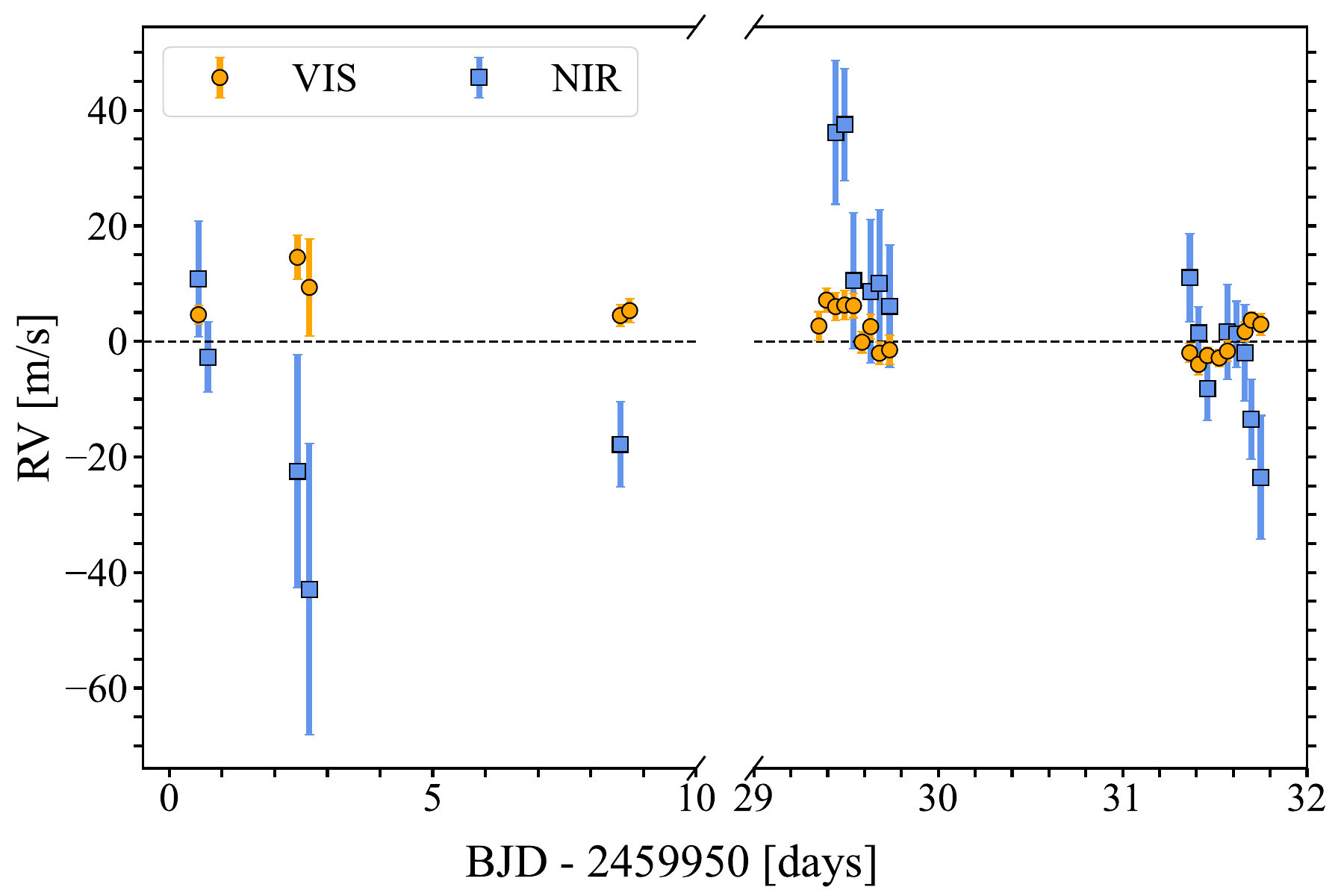}
    \caption{Wolf~327 CARMENES radial velocity measurements obtained using \texttt{SERVAL} for the visible (orange circles) and near-infrared (blue squares) channels.}
    \label{Fig:RVs_VISvsNIR}
\end{figure}

The \texttt{SERVAL} RV measurements for both channels are shown in Fig. \ref{Fig:RVs_VISvsNIR}. In this study, we used the RV data extracted using the VIS channel, as this dataset presented lower scatter than the NIR measurements (see \citealp{Bauer2020} for a comparison between CARMENES VIS and NIR RVs). The CARMENES VIS channel velocity measurements and activity indicators used in this study can be found in Tables \ref{Tab:CARMENES_RVs} and \ref{Tab:CARMENES_Activ}, respectively.

\section{Stellar properties}
\label{Sec:StarProper}

\begin{table*}
\centering
\small
\caption{Stellar parameters of Wolf~327.} \label{Tab:Star}
\begin{tabular}{lcr}
\hline\hline
\noalign{\smallskip}
Parameter                               & Value                 & Reference \\ 
\hline
\noalign{\smallskip}
\multicolumn{3}{c}{Identifiers}\\
\noalign{\smallskip}
Wolf                            & 327                           &  {\citet{Wolf1919}}.    \\
LSPM                            & J0953+3534                    & {\citet{Lepine2005}}      \\    
Karmn                           & J09535+355                    & {\citet{2016csss.confE.148C}}      \\    
TOI                             & 5747                          & TESS Alerts      \\  
TIC                             & 4918918                       & {\citet{Stassun2018}}  \\
{\it Gaia} DR3                  & 796185407950360192            & {\citet{GaiaDR3}}      \\
2MASS                           & J09533093+3534171             & {\citet{2MASS}}        \\
  
\noalign{\smallskip}
\multicolumn{3}{c}{Coordinates and spectral type}\\
\noalign{\smallskip}
$\alpha$ (ICRS, J2000)                        & 09:53:30.90           & {\it Gaia} DR3     \\
$\delta$ (ICRS, J2000)                        & +35:34:16.7          & {\it Gaia} DR3     \\
Spectral type                           & M2.5\,V               & {This work}        \\
\noalign{\smallskip}
\multicolumn{3}{c}{Photometry}\\
\noalign{\smallskip}

$B$ [mag]                               & $14.586 \pm 0.070$      & UCAC4      \\
$G_{BP}$ [mag]                          & $13.247 \pm 0.003$      & {\it Gaia} DR3 \\
$V$ [mag]                               & $13.018 \pm 0.062$      & UCAC4      \\
$G$ [mag]                               & $11.943 \pm 0.003$      & {\it Gaia} DR3 \\
$G_{RP}$ [mag]                          & $10.811 \pm 0.003$      & {\it Gaia} DR3 \\
$J$ [mag]                               & $9.308 \pm 0.022$       & 2MASS      \\
$H$ [mag]                               & $8.682 \pm 0.021$       & 2MASS      \\
$K_s$ [mag]                             & $8.435 \pm 0.017$       & 2MASS      \\
$W1$ [mag]                              & $8.320 \pm 0.023$       & AllWISE    \\
$W2$ [mag]                              & $8.211 \pm 0.020$       & AllWISE    \\
$W3$ [mag]                              & $8.104 \pm 0.021$       & AllWISE    \\
$W4$ [mag]                              & $7.838 \pm 0.165$       & AllWISE     \\
\noalign{\smallskip}
\multicolumn{3}{c}{Parallax and kinematics}\\
\noalign{\smallskip}
$\pi$ [mas]                             & $35.00\pm0.03$       & {\it Gaia} DR3             \\
$d$ [pc]                                & $28.56^{+0.02}_{-0.03}$ & {\citet{Bailer-Jones2021}}             \\
$\mu_{\alpha}\cos\delta$ [$\mathrm{mas\,yr^{-1}}$]  & $-87.76 \pm 0.02$ & {\it Gaia} DR3          \\
$\mu_{\delta}$ [$\mathrm{mas\,yr^{-1}}$]            & $-177.14 \pm 0.02$ & {\it Gaia} DR3          \\
$\gamma$ [$\mathrm{km\,s^{-1}}]$        & $9.39 \pm 0.42$  & {\it Gaia} DR3  \\
$U$ [$\mathrm{km\,s^{-1}}]$             & $-11.26 \pm 0.25$   & {This work}  \\
$V$ [$\mathrm{km\,s^{-1}}]$             & $-26.04 \pm 0.04$   & {This work}  \\
$W$ [$\mathrm{km\,s^{-1}}]$             & $-0.10 \pm 0.32$  & {This work}   \\
\noalign{\smallskip}
\multicolumn{3}{c}{Photospheric parameters}\\
\noalign{\smallskip}
$T_{\mathrm{eff}}$ [K]                      & $3542 \pm 70$         & {This work}   \\
$\log g$ [cm/s$^2$]                         & $4.87 \pm 0.05$       & {This work}   \\
{[Fe/H]} dex                                & $-0.17 \pm 0.04$      & {This work}   \\
$v\sin i$ [km/s]                            & $< 2$ & {This work}   \\
\noalign{\smallskip}
\multicolumn{3}{c}{Physical parameters}\\
\noalign{\smallskip}
$M_\star$ [$M_{\odot}$]                       & $0.405 \pm 0.019$     & {This work}       \\
$R_\star$ [$R_{\odot}$]                       & $0.406 \pm 0.015$     & {This work}       \\
$L$ [$10^{-4}\,L_\odot$]                & $234.5 \pm 1.2$       & {This work}       \\
$P_{\rm rot}$ [d] & $44.4 \pm 0.4$ & {This work} \\
Age [Gyr] & $4.1^{+3.2}_{-2.4}$ &  {This work} \\
\noalign{\smallskip}
\hline
\end{tabular}
\tablebib{
    {\it Gaia} DR3: \citet{GaiaDR3};
    UCAC: \citet{UCAC4};
    2MASS: \citet{2MASS};
    AllWISE: \citet{CutriWISECat2014}.
}
\end{table*}

\subsection{Stellar parameters}
The coordinates, photometric magnitudes, astrometric values, and stellar parameters of Wolf~327 are listed in Table \ref{Tab:Star}. The stellar parameters were computed using a combined CARMENES stellar spectrum. The effective temperature ($T_{\rm eff}$), surface gravity ($\log g$), and iron abundance ([Fe/H]) of the star were derived using \texttt{SteParSyn}\footnote{\url{https://github.com/hmtabernero/SteParSyn/}} code \citep{Tabernero2022}. \texttt{SteParSyn} makes use of the line list (for the visible and near-infrared wavelength ranges available in CARMENES data) and model grid described by \cite{Marfil2021}. The error in $T_{\rm eff}$ is the quadratic sum of the measurement error given by \texttt{SteParSyn} (8\,K) and the weighted average of the spreads of the differences to literature values (69\,K) given in \cite{Marfil2021}. The target's spectral type was determined with $\pm$0.5\,dex accuracy from the color-, absolute magnitude-, and luminosity-spectral type relations of \cite{Cifuentes2020}. The stellar mass ($M_\star$) and radius ($R_\star$) were determined following \cite{Schweitzer2019}, and the stellar luminosity ($L$) was computed following \cite{Cifuentes2020}. We determine that Wolf~327 is an M2.5 V dwarf with an effective temperature of $T_{\rm eff} = 3542 \pm 70$\,K, and a stellar mass and radius of $M_\star = 0.405 \pm 0.019 \; M_\odot$ and $R_\star = 0.406 \pm 0.015 \; R_\odot$, respectively. 

We searched for common proper-motion companions to Wolf\,327 in the \textit{Gaia} DR3 catalog up to a radius of 1$^{\circ}$. We imposed a restriction on the parallax with a range of 35$\pm$5\,mas, bracketing the parallax of Wolf\,327 (see Table \ref{Tab:Star}). The query returned only Wolf\,327, which thus does not seem to share motion {\rm and} distance with any other star within one degree down to the depth of \textit{Gaia}. At the distance of Wolf\,327 ($d=28.57\pm1.02$\,pc), \textit{Gaia} is sensitive to all stellar companions and even the highest-mass brown dwarfs \citep{Reyle2021, Lodieu2019}.

The galactocentric velocities $(U,V,W)$ of the star presented in Table \ref{Tab:Star} were computed following \cite{CortesContreras2017}, using \textit{Gaia} DR3 astrometry and adopting a right-handed system with no solar motion correction. The derived Galactic velocities indicate that Wolf~327 belongs to the thin disk \citep{Montes2001}, suggesting an age $> 1$\,Gyr. According to \texttt{BANYAN $\Sigma$}\footnote{\url{https://www.exoplanetes.umontreal.ca/banyan/banyansigma.php}} \citep{Gagne2018}, the $U,V,W$ values suggest that Wolf~327 does not belong to any young moving group, a finding supporting a relatively old age for this star.

\subsection{Stellar rotation from seeing-limited photometry}
\label{Sec:StellarRotation}
We used the long-term photometric $V$-band data from ASAS-SN and the $R$-band data from TJO to search for periodic flux variations attributable to the rotation period of the star. We attempted to fit the available datasets jointly using a unified modeling approach. However, when adopting a common modeling procedure, the final model for the TJO photometry exhibited clear signs of overfitting. Consequently, we decided to rely solely on the ASAS-SN dataset to establish the rotation period of the star. In this section, we describe our adopted results for the ASAS-SN dataset, while our results for TJO are presented in Section \ref{Sec:Appendix_TJO_photometry}.

To model the flux variations, we used a combination of a linear function in addition to a periodic term using Gaussian processes \citep[GPs; e.g.,][]{Rasmussen2006,Gibson2012} of the form
 \begin{equation}
     L(t) = z_{\mathrm{pt}} + \alpha t + k_{ij\; \mathrm{QP}}
  \label{Eq:ASASSN-lightcurve}
 ,\end{equation}
where $z_{\mathrm{pt}}$ and $\alpha$ are the zero point and slope of the linear function, respectively, and $k_{ij\; \mathrm{QP}}$ is a GP periodic kernel. For the periodic term, we used the implementation provided by the GP package \texttt{George} \citep{Ambikasaran2015} and chose the quasi-periodic kernel described by
 \begin{equation}
    k_{ij\; \mathrm{QP}} = A \exp \left[ \frac{-(|t_i-t_j|)^2}{2 l^2}
 - \Gamma^2 \sin^2\left( \frac{\pi |t_i-t_j|}{P_{\rm rot}} \right) \right]
\label{Eq:Prot_GP}
,\end{equation}
where $|t_i-t_j|$ is the difference between two epochs or observations, $A$ is the variance, $l$ is the evolution timescale, $\Gamma$ is the scale factor and $P_{\rm rot}$ is the period of the sinusoidal variation, that is, the rotational period of star \citep[for a discussion of the physical interpretation of these terms see][]{Nicholson2022}. We also added a jitter term ($\sigma_{\mathrm{phot\; jitter}}$) as a free parameter to fit the white noise component present in the time series.

The fitting procedure was performed in two steps: first we optimized a posterior probability function using \texttt{PyDE}\footnote{\url{https://github.com/hpparvi/PyDE}} and then we used the optimal set of parameters to start exploring the parameter space using \texttt{emcee} \citep{ForemanMackey2013}. We let \texttt{emcee} run for 3\,000 iterations as a burn-in period and then ran the main Markov chain Monte Carlo (MCMC) procedure for 10\,000 iterations. The final values of the fitted parameters were obtained by computing the median and 1$\sigma$ percentiles of the posterior distributions. Table \ref{Tab:Prot_parameters} presents the values and 1$\sigma$ uncertainties of our fit.

\begin{figure*}[]
   \centering
   \includegraphics[width=\hsize]{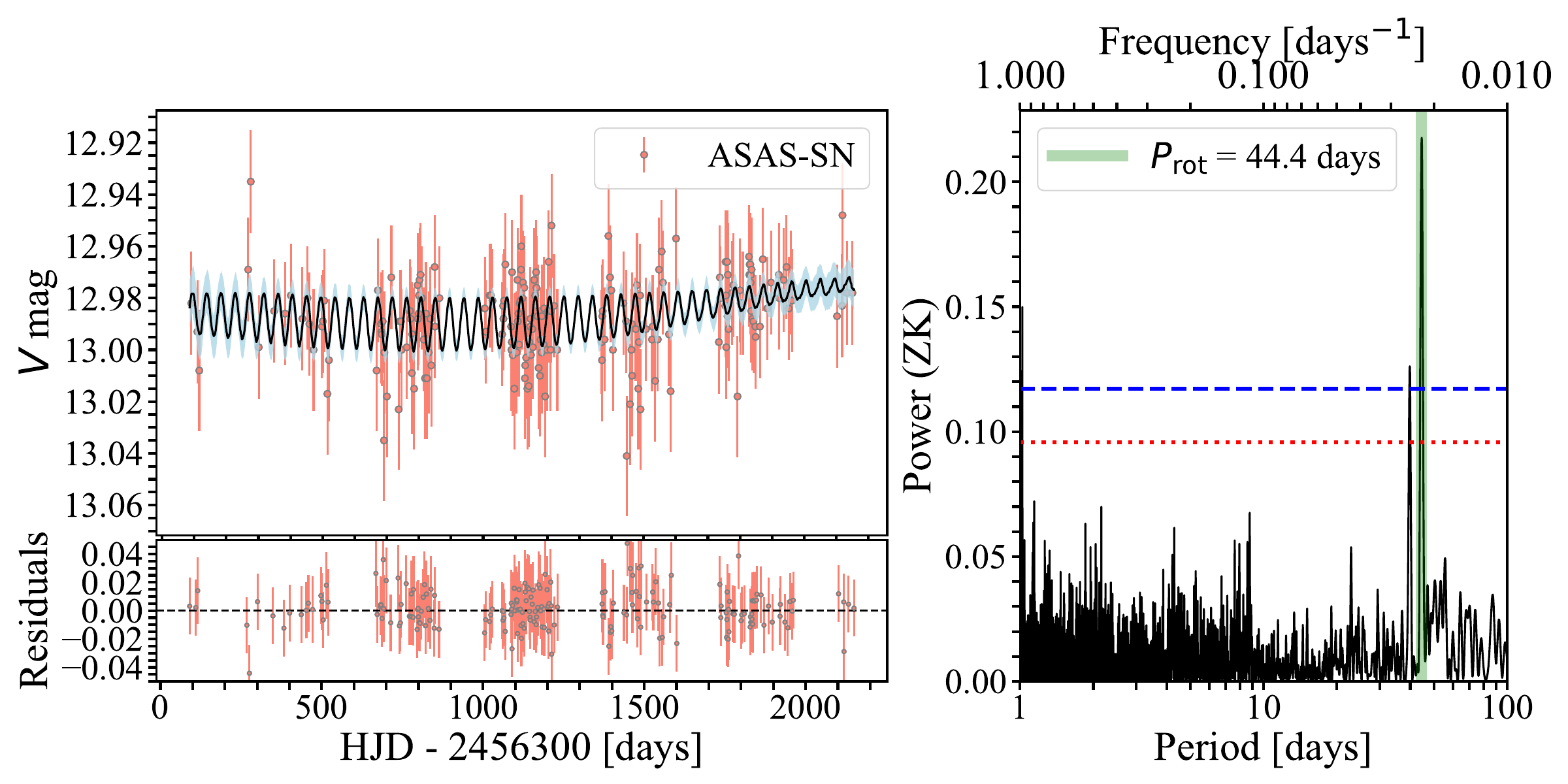}
   \caption{Long-term photometric archival data of Wolf~327 from ASAS-SN ($V$ band). \textit{Left panel:} photometric time-series and best-fitted model computed using GP adopting a quasi-periodic kernel (black line). The 1$\sigma$ uncertainty regions of the fit are shown in light blue. The residuals of the fit are shown in the bottom panel. \textit{Right panel:} GLS periodogram \citep{Zechmeister2009} for ASAS-SN $V$-band photometry. The vertical green line shows the fitted rotational period of the star. The horizontal lines represent the false alarm probability (FAP) levels of 10\% (red dotted line) and 1\% (blue dashed line).}
    \label{Fig:Wolf327_Prot}
\end{figure*}

Figure \ref{Fig:Wolf327_Prot} shows the photometric ASAS-SN $V$-band time series and its periodogram. The left panel shows the median model (black line) computed using the posterior distributions of the fitted parameters; the blue-shaded area shows 1$\sigma$ uncertainty of the model. The right panel shows the Generalized Lomb-Scargle (GLS) periodogram computed using the \texttt{python} implementation by \citet{Zechmeister2009}.\footnote{\url{https://github.com/mzechmeister/GLS}} The root-mean-square of the residuals of the fit is $rms = 0.013$\,mag.

The periodogram for the ASAS-SN observations presents a significant peak around 44\,d, which we attribute to the stellar rotation of the star. With our fitting procedure we find a stellar rotational period of $P_{\rm rot} = 44.4 \pm 0.4$ d, which will be our final adopted value. This value is in agreement with the results found for the TJO data set ($P_{\mathrm{rot}} = 42.1^{+3.5}_{-5.8} \; d$, see Sec. \ref{Sec:Appendix_TJO_photometry}). Finally, we used the age--rotation period relation described in \cite{Angus2015} to get an estimate of the age of the star. Using our rotation period value (including uncertainties) and equation 1 of \cite{Angus2015}, we find an age of $4.1^{+3.2}_{-2.4}$\,Gyr. Considering the uncertainties, this value is in agreement with the estimate obtained using the age--rotation relation for M dwarfs derived by \cite{EngleGuinan2018}; with this relation we obtain an age of $2.7 \pm 1.1$\,Gyr.

\begin{table}[t]
\centering
\caption{Prior functions and fitted parameters values for the ASAS-SN $V$-band photometry of Wolf~327.}
\label{Tab:Prot_parameters}
\renewcommand{\arraystretch}{1.3}
\begin{tabular}{l c c}
\hline 
\hline
\noalign{\smallskip}
Parameter & Prior & Value \\
\noalign{\smallskip}
\hline
\noalign{\smallskip}

$z_{\mathrm{pt}}$ [mag] & $\mathcal{U}(-20, 20)$ & $12.98 \pm 0.02$ \\
$\alpha \; (\times 10^{-6})$ [mag/d] & $\mathcal{U}(-50, 50)$ & $6.7^{+17.3}_{-16.2}$ \\
$A$ [mmag] & $\mathcal{U}(10^{-6}, 15)$ & $0.53^{+2.5}_{-0.4}$ \\
$l$ [d] & $\mathcal{U}(1.0, 1000)$ & $841^{+115}_{-200}$ \\
$\Gamma$ & $\mathcal{U}(10^{-6}, 100)$ & $0.44^{+0.51}_{-0.27}$ \\
$P_{\mathrm{rot}}$ [d] & $\mathcal{U}(1.0, 100)$ & $44.4 \pm 0.4$ \\
$\sigma_{\mathrm{phot\; jitter}}$ [mmag] & $\mathcal{U}(10^{-6}, 10)$ & $1.3^{+1.4}_{-0.9}$ \\
\noalign{\smallskip}
\hline
\end{tabular}
\renewcommand{\arraystretch}{1}
\tablefoot{$\mathcal{U}$ represent a uniform prior function. The slope of the linear function ($\alpha$) was computed relative to the mid-time of the time series $T_{\mathrm{mid}} = 2457420$\,HJD.}
\end{table}

\subsection{Stellar activity indicators}
We investigated the spectral line diagnostics of stellar activity across both the VIS and NIR channels of CARMENES. In the CARMENES wavelength range we computed a total of sixteen activity indices following the methods described in \cite{Schofer2019} and \cite{Jeffers2018}. Because the spectra of M dwarfs do not exhibit extensive continuum regions, the spectral line activity indices were computed using the pseudo-equivalent width (pEW) of the lines by integrating the fluxes in regions centered at the position of the line and in two nearby regions (see Table 1 of \citealp{Schofer2019}). When dealing with the absorption bands of TiO and VO present in the spectrum, the pEW proves ineffective as a measure for quantifying strength. This inadequacy arises due to the potential extension of these bands across multiple spectral orders, where they can become blended with other spectral features. Instead of using pEWs, the TiO and VO activity levels were computed by measuring the mean fluxes in two regions and taking the ratio between these two quantities (see Table 3 of \citealp{Schofer2019}). The same approach was employed for the iron hydride (FeH) Wing-Ford band \citep{WingFord1969,Schiavon1997,Schofer2021}. The regions where the mean fluxes were computed are defined by 989.8 – 990.7\,nm and 988.7 – 989.6\,nm (see Table 2.3 of \citealp{Schofer2021}). The photospheric and chromospheric line activity indices studied here were: $\log (L_{\mathrm{H}\alpha}/L_\mathrm{bol})$, pEW(H$\alpha$), He D$_3$, Na D, Ca IRT-a,-b,-c, He I 10830, Pa$\beta$, CaH$_2$, CaH$_3$, TiO 7050, TiO 8430, TiO 8860, VO 7436, VO 7942, and FeH Wing-Ford (see Table \ref{Tab:ActivIndicators}).

In addition to the absorption line indicators, we also investigated the RV chromatic (CRX) and the differential line width (dLW) indices. These indicators serve the purpose of monitoring the stellar activity of the star and assessing its impact on the RV measurements. The CRX index is the slope of the radial velocity (computed in each echelle spectral order) as a function of the logarithm of the wavelength \citep{Zechmeister2018}. The dLW index is used as a proxy to track differential changes in the absorption line widths of the spectrum. This index is computed by comparing the second derivative of the high signal-to-noise spectrum template to the residuals from the fit for the best RVs. This comparison is performed in each echelle spectral order and then averaged taking into account the uncertainties of the spectra \citep{Zechmeister2018}. As is the case for the RVs, we use only the CRX and dLW values derived using the CARMENES VIS channel. Considering all the stellar activity indices, we studied a total of twenty activity diagnostic parameters.

Our results show that Wolf~327 is a relatively stable and H$\alpha$ inactive star. Stars with positive pEW(H$\alpha$) values have H$\alpha$ lines in absorption in contrast to H$\alpha$-active stars with H$\alpha$ in emission and negative pEW(H$\alpha$) values. The low activity level is consistent with the $\sim$44 day rotation period of Wolf~327. 

We also investigated whether there was a statistical correlation between the twenty activity diagnostics and the RV values using Pearson’s $r$, as previously described by \cite{Jeffers2020}. None of the twenty diagnostics had a statistically significant correlation with the RV which is defined as Pearson’s $r$ having a value $r > |0.7|$. Given the low number of RV measurements, any further statistical analysis would be misleading.

\begin{table}[t]
\centering
\caption{Wolf~327 activity indicators computed from a coadded CARMENES VIS and NIR channel spectrum.}
\label{Tab:ActivIndicators}
\begin{tabular}{l c}
\hline 
\hline
\noalign{\smallskip}
Indicator & Value \\
\noalign{\smallskip}
\hline
\noalign{\smallskip}

pEW($\mathrm{H}\alpha$) & $0.0256 \pm 0.0087$ \\
$\log(L_{{\rm H}\alpha}/L_{\mathrm{bol}})$ & Inactive \\
He D$_3$ & $0.0201 \pm 0.0083$ \\
Na D & $-0.4883 \pm 0.0203$ \\
Ca IRT a & $0.0320 \pm 0.0062$ \\
Ca IRT b & $0.0302 \pm 0.0048$ \\
Ca IRT c & $0.0282 \pm 0.0031$ \\
Fe 8691 & $0.0132 \pm 0.0031$ \\
He 10833 & $0.0073 \pm 0.0006$ \\
Pa$\beta$ & $0.0002 \pm 0.0007$ \\
CaH$_2$ & $0.9197 \pm 0.0008$ \\
CaH$_3$ & $0.7644 \pm 0.0004$ \\
TiO 7050 & $0.6506 \pm 0.0003$ \\
TiO 8430 & $0.8407 \pm 0.0007$ \\
TiO 8860 & $0.9125 \pm 0.0006$ \\
VO 7436 & $0.9237 \pm 0.0008$ \\
VO 7942 & $0.9612 \pm 0.0005$ \\
FeH Wing-Ford & $0.9696 \pm 0.0021$ \\

\noalign{\smallskip}
\hline
\end{tabular}
\end{table}

\section{Analysis and results}
\label{Sec:AnalysisResults}

\subsection{Radial velocity and activity indices periodogram analysis}
\label{Sec:RVsPeriodo}
\begin{figure}
    \centering
    \includegraphics[width=\hsize]{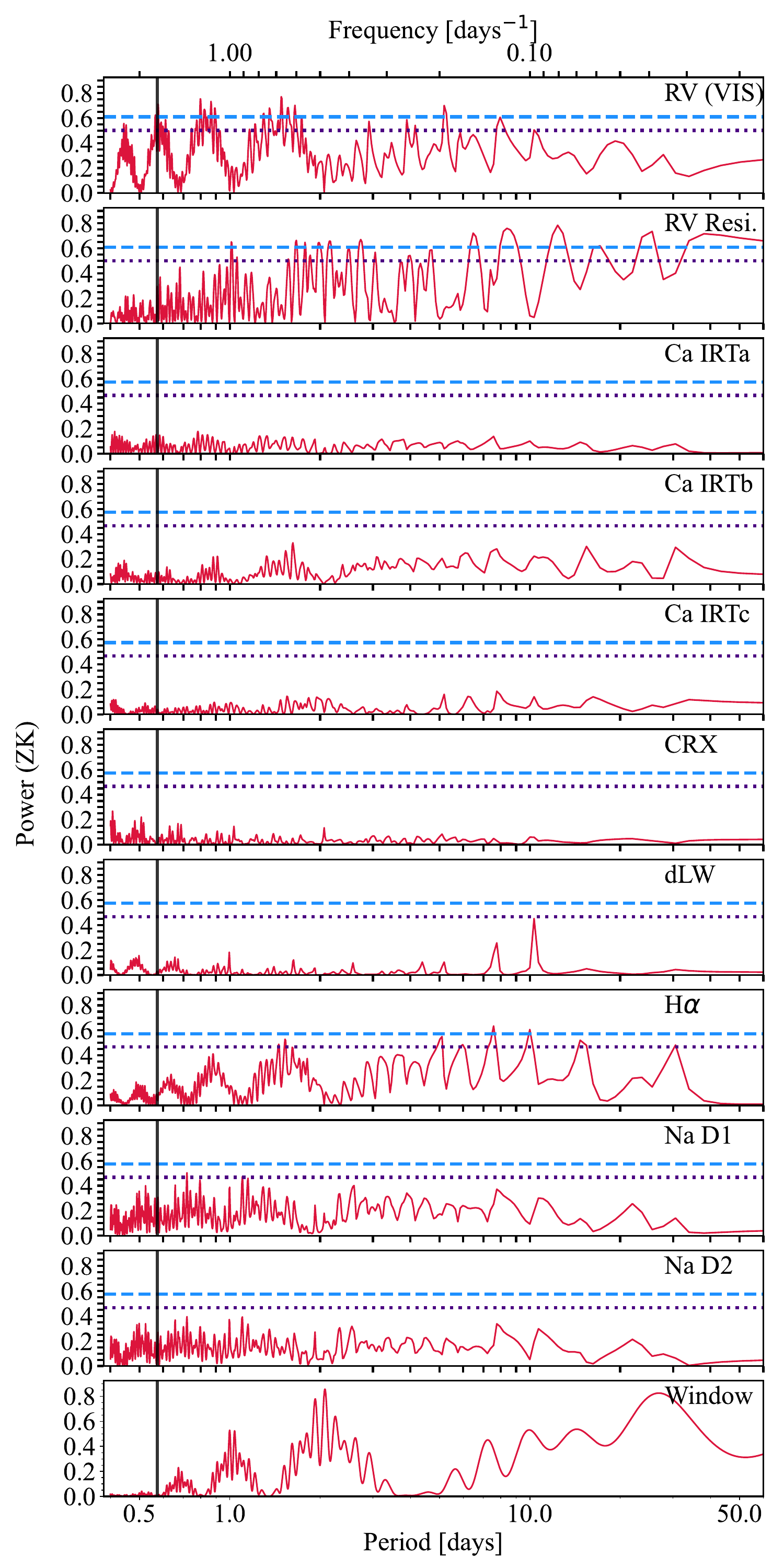}
    \caption{GLS periodograms of the CARMENES RV measurements taken with the optical channel, RV residuals (i.e., linear trend and planet signal removed), spectral activity indices, and window function. The horizontal lines represent the FAP levels of 10\% (purple dotted line) and 1\% (blue dash line). The vertical black line marks the orbital period of Wolf~327b ($P=0.5734$\,d).}
    \label{Fig:RVActiv_Periodogram}
\end{figure}

We computed the GLS periodogram of the CARMENES radial-velocity and activity-indicator measurements for Wolf~327. Figure~\ref{Fig:RVActiv_Periodogram} shows the periodograms for the CARMENES RV measurements (VIS channel), the RVs after subtracting a linear trend and the planetary signal (see Sect. \ref{Sec:JointFit}), and the activity indices provided by \texttt{SERVAL}, the vertical black line marks the orbital period of the transiting candidate around Wolf~327.

The RV periodogram shows a broad peak consistent with the period of the transiting candidate detected by TESS. The peak is slightly above the 1\% false alarm probability (FAP) level, although this peak is not the one with the highest power in the periodogram. The RV signature of the rotation period of the star is not present in any of the periodograms, probably because the RV measurements were taken with a time baseline shorter than the rotation period of the star (31\,d versus $P_{\rm rot} = 44.4 \pm 0.4$\,d from photometric measurements) and the observation strategy, as the majority of the RV data acquisition was made during two full nights of continuous observations.

\subsection{TESS, ground-based light curves, and CARMENES data}
\label{Sec:JointFit}
To derive the planetary orbital parameters we decided to use TESS, ground-based light curves, and CARMENES RV measurements to perform a joint fit of the data. We use the same method as described in previous papers \citep[e.g.,][]{Murgas2022,Murgas2023}. To model the transits we use \texttt{PyTransit}\footnote{\url{https://github.com/hpparvi/PyTransit}} \citep{Parviainen2015} and we adopted a quadratic limb darkening (LD) law. The fitted LD coefficients were compared through a likelihood function to the predicted values computed by \texttt{LDTK}\footnote{\url{https://github.com/hpparvi/ldtk}} \citep{Parviainen2015b} using the stellar parameters presented in Table \ref{Tab:Star}. During the fitting process we used the LD parameterization of \cite{Kipping2013}. The CARMENES radial velocity time series was modeled using \texttt{RadVel}\footnote{\url{https://github.com/California-Planet-Search/radvel}} \citep{Fulton2018}.

For the analysis we selected only ground-based observations that covered a full transit of the planet candidate and presented a detection of the transit event (for details see Appendix \ref{Sec:Appendix_GroundBasedLC}). In the end we used the observations taken on 2022 November 19, 2022 December 2, and 2022 December 12 by LCOGT 1 m telescopes; and the 2023 January 19 and 2023 April 25 time series taken with LCOGT 2 m telescope using the MuSCAT3 instrument. We used ground-based light curves that were corrected by observational and instrumental effects (i.e.,\ detrended light curves) provided by the ExoFOP team.

To model the systematic effects of unknown origin present in TESS and ground-based photometry we used GPs. For all the light curves analyzed here we adopted the commonly used Mat\'ern 3/2 kernel:
\begin{equation}
    k_{ij} = c^2_k \left( 1 + \frac{\sqrt{3} |t_i-t_j|}{\tau_k}\right) \exp\left(-\frac{\sqrt{3} |t_i-t_j|}{\tau_k}\right)
\label{Eq:TESS_GPKernel}
,\end{equation}
where $|t_i-t_j|$ is the time between epochs in the series, and the hyperparameters, $c_k$ and $\tau_k$, were allowed to be free, with $k$ indicating the different time series (i.e.\ individual TESS sectors, ground-based observations of transit events). 

We performed an initial test, fitting TESS data and CARMENES measurements simultaneously using RV models with and without GPs, and a zero-point offset for each observing night. We found that all the methodologies tested deliver similar results in the determination of the radial velocity semi-amplitude (see Appendix \ref{Sec:Appendix_JointFitResults}, Fig.\ \ref{Fig:RVAmp_Test_GPs_vs_FloatingChunk}). For the case in which we included GPs in the RVs to model the systematic noise present in the measurements, we adopted a Gaussian kernel of the form
\begin{equation}
    k_{ij\; \mathrm{RV}} = c^2_{rv} \exp \left(  - \frac{ (t_i-t_j)^2}{\tau^2_{rv}} \right)
\label{Eq:RV_GPKernel}
,\end{equation}
where $t_i-t_j$ is the time between epochs in the series, and the hyperparameters, $c_{rv}$ and $\tau_{rv}$, were set free. We decided to use this kernel since it has few free parameters and assumes that points closer in time will be more correlated than observations separated by a long time baseline. With this approach we expect to model the effect on the RVs of transient stellar activity events on the star. The use of a more complex kernel, such as the popular quasi-periodic kernel, is not warranted since the RV periodogram did not show any evidence of a strong correlation with the rotation period of the star. 

To model the planetary transits and RV variations we set as free parameters the planet-to-star radius ratio ($R_\mathrm{p}/R_\star$), the LD coefficients ($q_1$, $q_2$ following \citealp{Kipping2013}), the central time of the transit ($\mathrm{T}_{\mathrm{c}}$), the planetary orbital period ($P$), the stellar density ($\rho_\star$), the transit impact parameter ($\mathrm{b}$), the RV semi-amplitude ($K_\mathrm{RV}$), the host star systemic velocity ($\gamma_0$), the long-term linear trend of the radial velocity variation ($\Dot{\gamma}$) and the RV jitter ($\sigma_{\mathrm{RV\; jitter}}$). Owing to the short orbital period of the candidate, we decided to assume a circular orbit and fix the eccentricity to 0. As shown in Fig.\ \ref{Fig:Wolf327b_TESS_LC}, TESS light curves show deep eclipses caused by the nearby star TIC~4918919 (see Sect. \ref{Sec:SpaceObs}); we model the contribution of this EB with its own light curve model by setting as free parameters the transit depth, period, central time of transit, stellar density, and impact parameter. The LD coefficients for TIC~4918919 were fixed to the values computed with \texttt{LDTK} using the TESS Input Catalog \citep{Stassun2018,Stassun2019} stellar parameter values for this star.

The fitting procedure started with the optimization of a posterior probability function with \texttt{PyDE}. After the optimization finished, we used the optimal parameter vector as a start of an MCMC procedure using \texttt{emcee} to explore the parameter space. We ran the MCMC using a number of chains that were at least 5.5 times the number of free parameters for 35\,000 iterations as a burn-in stage to ensure convergence. We then ran the main MCMC for another 15\,000 iterations. We computed the final values of the fitted parameters using the median and 1$\sigma$ limits of the posterior distributions.

We compared the results of different modeling approaches for the joint fit of all the datasets available (i.e.\ TESS, ground-based photometry; CARMENES RVs) through the Bayesian information criterion metric \citep[BIC,][]{Schwarz1978}. Because of the photometric variability present in TESS and ground-based light curves, we decided to keep the use of GPs for these datasets. In the case of the RV measurements we tested models without and with GPs. We computed the BIC metric for three circular orbit models using different modeling approaches for the RVs: Model 1) RVs without GPs, Model 2) RVs including GPs, and Model 3) RVs using the zero-point offset method (no GPs). The computed BIC differences between the models were $\Delta \mathrm{ BIC_{M_1-M_2} } = -10.4$ and $\Delta \mathrm{ BIC_{M_1-M_3} } = -74.4$, meaning that Model 1 presented the lowest BIC value. Therefore, according to the BIC metric, the best model was the simplest one, i.e.\ transit analysis including GPs for TESS and ground-based observations and no GPs for the RV measurements. Hence, we adopted the parameter values of Model 1 as our final results. Using this approach we had a total of 50 free parameters with which to model the planet candidate (transits and RVs), the eclipsing binary TIC~4918919 in TESS light curves, and systematic effects.

\begin{figure}
    \centering
    \includegraphics[width=\hsize]{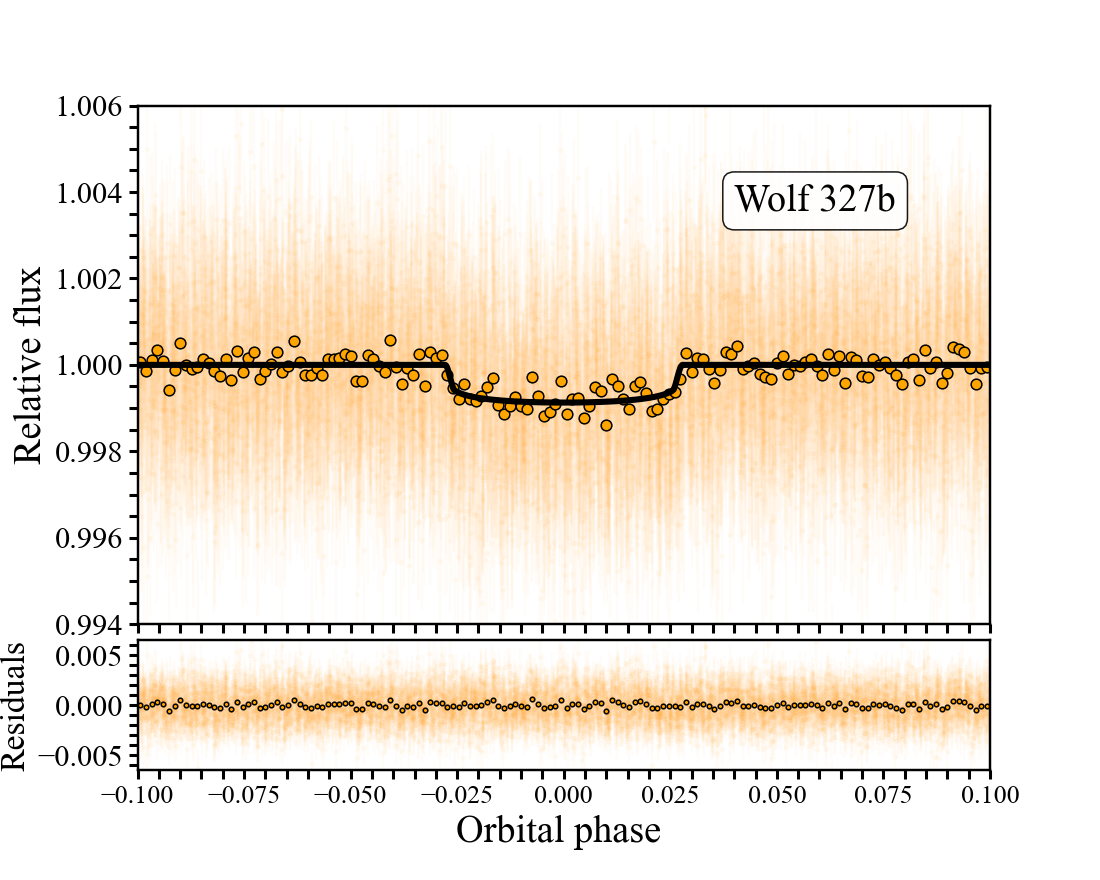}
    \caption{TESS phase-folded light curve of Wolf~327b after subtracting the photometric variations from the time series (top panel). The best-fit transit model is shown in black, the circles are TESS binned observations. \textit{Bottom panel}: residuals of the fit.}
    \label{Fig:Wolf327b_TESS_LC_phase}
\end{figure}

\begin{figure*}
   \centering
   \includegraphics[width=\hsize]{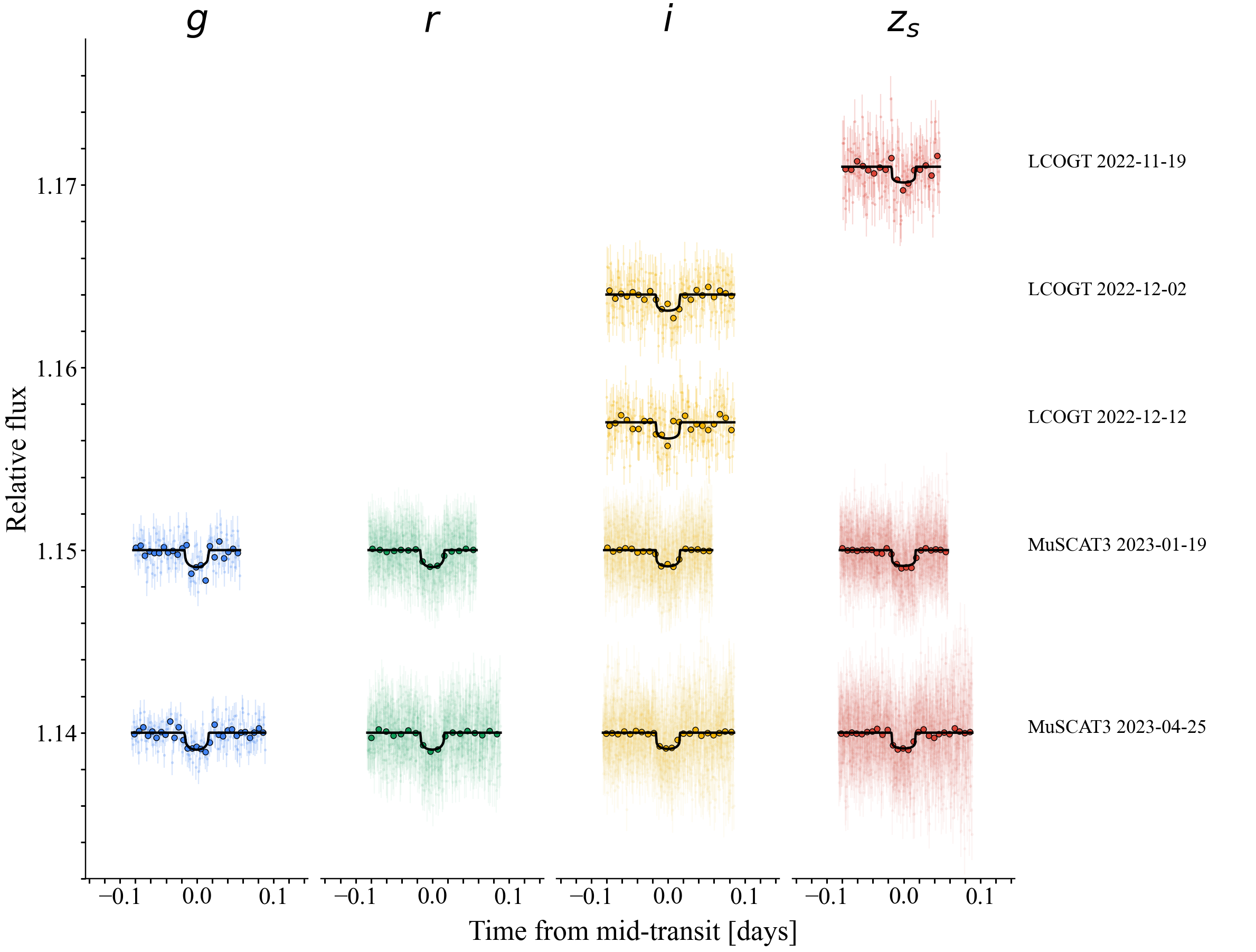}
   \caption{Ground-based individual transit observations of Wolf~327b used in the joint fit, systematic effects were removed from the light curves. The points show the individual observations, the circles represent binned data points, and the best-fit model is shown by the black line.}
    \label{Fig:GroundBased_LightCurves_JointFit}
\end{figure*}

\begin{figure}
    \centering
    \includegraphics[width=\hsize]{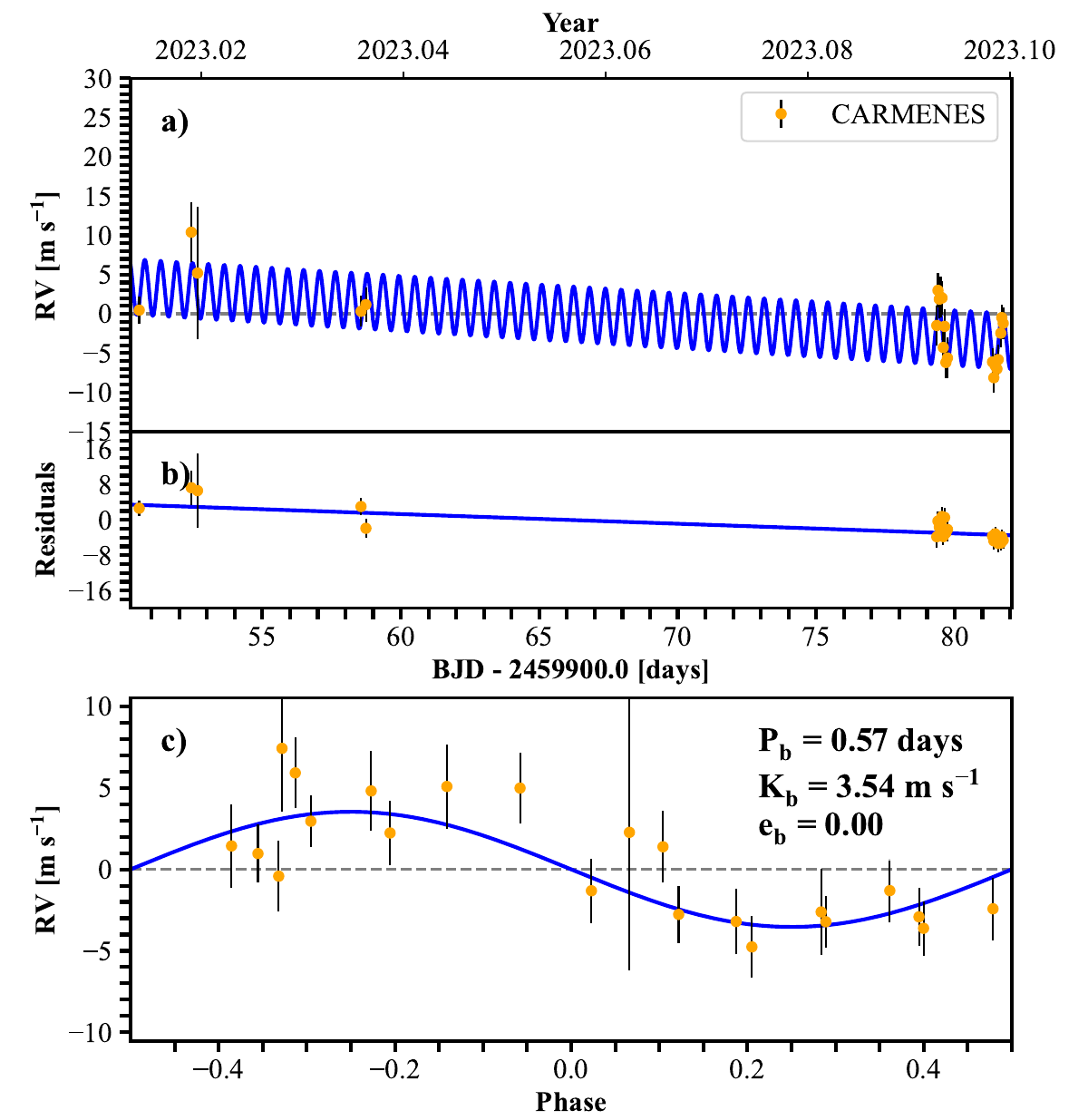}
    \caption{Wolf~327b radial velocity measurements taken with CARMENES. (a) RV time series and best-fitting model (blue line). The best-fitting model was computed using the median values of the posterior distribution of the fitted parameters. (b) Residuals of the fit after subtracting the single-planet Keplerian, the velocity linear trend seen here was included in the RV modeling. (c) Phase-folded RV measurements.}
    \label{Fig:Wolf327b_CARM_RV_phase}
\end{figure}

\begin{figure}
    \centering
    \includegraphics[width=\hsize]{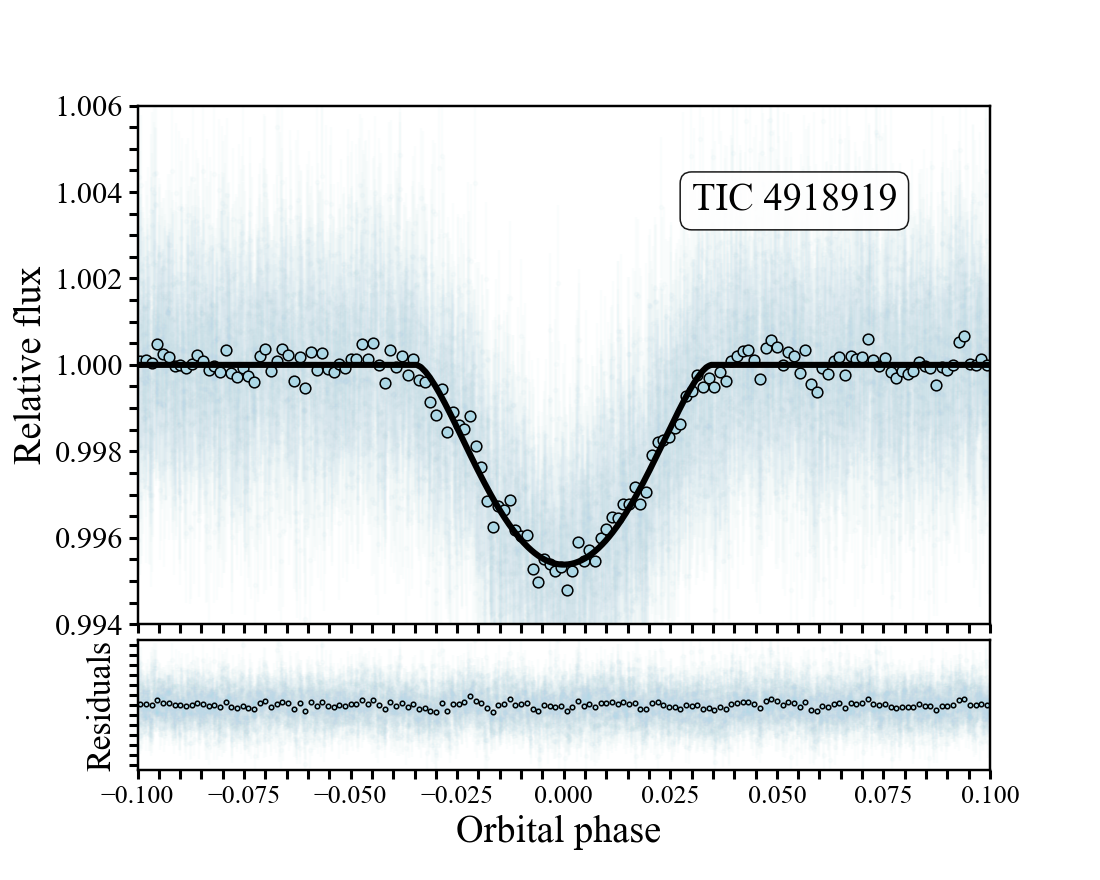}
    \caption{\textit{Top panel}: TESS phase-folded light curve of the eclipsing binary TIC~4918919. The photometric variations modeled by the GPs were removed from the time series. The best-fit transit model is shown in black, the circles are TESS binned observations. \textit{Bottom panel}: residuals of the fit.}
    \label{Fig:EB_TESS_LC_phase}
\end{figure}

Table \ref{Tab:Planet_parameters} presents the median and 1$\sigma$ uncertainty range for the fitted and derived parameters for Wolf~327b. In Appendix \ref{Sec:Appendix_JointFitResults}, Figure \ref{Fig:MCMC_CornerPlot} shows the posterior distributions of the transit and RV measurements fitted parameters (without the GP hyperparameters for easy viewing). Figures \ref{Fig:Wolf327b_TESS_LC_phase}, \ref{Fig:GroundBased_LightCurves_JointFit}, and \ref{Fig:Wolf327b_CARM_RV_phase} present the best-fitting model for TESS and ground-based light curves, and RV measurements, respectively. Figure \ref{Fig:EB_TESS_LC_phase} shows the phase-folded TESS data and best-fitting model for the eclipsing binary TIC~4918919. The radial velocity residuals (see Fig.\ \ref{Fig:Wolf327b_CARM_RV_phase} middle panel) shows a linear trend with time. This trend may be caused by another massive object in the system; however, the relatively short baseline of the RV observations (31\,d) does not allow us to put limits on the mass of the object. The potential presence of other planets in Wolf~327 system is in agreement with what is known of the USP population \citep{SanchisOjeda2014, Winn2018}; however, more RV observations are needed to confirm the presence of more objects in the system and rule out certain stellar activity induced signal that may be the caused of this linear trend.

Using our fitted values and the stellar parameters presented in Table \ref{Tab:Star}, we find that Wolf~327b has a radius of $R_\mathrm{p} = 1.24 \pm 0.06 \; R_\oplus$ and a mass of $M_\mathrm{p} = 2.53 \pm 0.46 \; M_\oplus$. This planet orbits its host star with a period of $P = 0.57347$\,d (13.76\,h) and has an equilibrium temperature of $T_\mathrm{eq} = 996 \pm 22$\,K. 

\begin{table*}[t]
\centering
\caption{Prior functions, fitted, and derived parameters of Wolf~327b. The fitted GP hyperparameters were left out on purpose for easy viewing.}
\label{Tab:Planet_parameters}
\begin{tabular}{l c c}
\hline 
\hline
\noalign{\smallskip}
Parameter & Prior & Value \\
\noalign{\smallskip}
\hline
\noalign{\smallskip}
\multicolumn{3}{c}{Fitted transit and orbital parameters} \\
\noalign{\smallskip}

$\rho_*$ [g cm$^{-3}$] & $\mathcal{N}(8.5,1.08)$ & $8.54 \pm 0.30$ \\
$R_\mathrm{p}/R_{*}$ & $\mathcal{U}(0.0015, 0.10)$ & $0.0280 \pm 0.0007$ \\
$T_{c} - 2\,457\,000$  [BJD] & $\mathcal{U}(2252.8167,2253.1167)$ & $2252.9811 \pm 0.0003$ \\
$P$ [d] & $\mathcal{U}(0.4734,0.6734)$ & $0.5734745 \pm 0.0000003$ \\
$b = a/R_{*}\cos(i)$ & $\mathcal{U}(0,1)$ & $0.47 \pm 0.03$ \\
$e$  & &  0 (fixed) \\
$K_\mathrm{RV}$ [m/s] & $\mathcal{U}(0,60)$ & $3.54^{+0.61}_{-0.62}$ \\[2pt]
$\gamma_0$ [m/s] & $\mathcal{U}(-22.2,27.8)$ & $4.15^{+0.65}_{-0.68}$ \\[2pt]
$\Dot{\gamma}$ [m/s d$^{-1}$] & $\mathcal{U}(-100,100)$  & $-0.22^{+0.04}_{-0.05}$ \\[2pt]
$\sigma_{RV}$ [m/s] & $\mathcal{U}(0.01, 12.0)$ & $0.69^{+0.71}_{-0.48}$ \\

\noalign{\smallskip}
\multicolumn{3}{c}{Derived orbital parameters} \\
\noalign{\smallskip}

$a/R_*$ & & $5.30 \pm 0.06$ \\
$i$ [deg] & & $84.89^{+0.41}_{-0.38}$ \\

\multicolumn{3}{c}{Derived planet parameters} \\
\noalign{\smallskip}

$R_\mathrm{p}$ [R$_{\oplus}$] & & $1.24 \pm 0.06$ \\ 
$M_\mathrm{p}$ [M$_{\oplus}$] & & $2.53 \pm 0.46$ \\
$\rho_\mathrm{p}$ [g cm$^{-3}$] & & $7.24 \pm 1.66$ \\
$g_\mathrm{p}$ [m s$^{-2}$] & & $16.0 \pm 3.3$ \\
$a$ [au] & & $0.0100 \pm 0.0004$ \\
$\langle F_{\rm p} \rangle$ [$10^3$\,W\,m$^{-2}$] & & $318.3 \pm 25.0$ \\
$S_{\rm p}$ [$S_\oplus$] & & $233.9 \pm 18.3$ \\
$T_{\rm eq}$ ($A_{\rm Bond} = 0.3$) [K] & & $996 \pm 22$ \\

\noalign{\smallskip}
\multicolumn{3}{c}{Fitted LD coefficients} \\
\noalign{\smallskip}

$q_{1\;TESS}$ & $\mathcal{U}(0,1)$ & $0.271 \pm 0.007$ \\
$q_{2\;TESS}$ & $\mathcal{U}(0,1)$ & $0.260 \pm 0.007$ \\
$q_{1\;g}$ & $\mathcal{U}(0,1)$ & $0.598 \pm 0.025$ \\
$q_{2\;g}$ & $\mathcal{U}(0,1)$ & $0.309 \pm 0.013$ \\
$q_{1\;r}$ & $\mathcal{U}(0,1)$ & $0.605 \pm 0.034$ \\
$q_{2\;r}$ & $\mathcal{U}(0,1)$ & $0.276 \pm 0.022$ \\
$q_{1\;i}$ & $\mathcal{U}(0,1)$ & $0.344 \pm 0.013$ \\
$q_{2\;i}$ & $\mathcal{U}(0,1)$ & $0.258 \pm 0.009$ \\
$q_{1\;z}$ & $\mathcal{U}(0,1)$ & $0.223 \pm 0.006$ \\
$q_{2\;z}$ & $\mathcal{U}(0,1)$ & $0.254 \pm 0.007$ \\

\noalign{\smallskip}
\multicolumn{3}{c}{Derived LD coefficients} \\
\noalign{\smallskip}

$u_{1\;TESS}$ & & $0.270 \pm 0.005$ \\
$u_{2\;TESS}$ & & $0.250 \pm 0.010$ \\
$u_{1\;g}$ & & $0.477 \pm 0.013$ \\
$u_{2\;g}$ & & $0.296 \pm 0.026$ \\
$u_{1\;r}$ & & $0.430 \pm 0.025$ \\
$u_{2\;r}$ & & $0.349 \pm 0.040$ \\
$u_{1\;i}$ & & $0.303 \pm 0.007$ \\
$u_{2\;i}$ & & $0.284 \pm 0.015$ \\
$u_{1\;z}$ & & $0.240 \pm 0.004$ \\
$u_{2\;z}$ & & $0.233 \pm 0.010$ \\

\noalign{\smallskip}
\hline
\end{tabular}
\tablefoot{$\mathcal{U}$, $\mathcal{N}$ represent uniform and normal prior functions respectively. $A_{\rm Bond}$ is the Bond albedo. The term $\Dot{\gamma}$ was computed relative to $T_{\mathrm{base}} = 2459966.0$\,BJD. }
\end{table*}

\section{Discussion}
\label{Sec:Discussion}
\subsection{Comparison with other exoplanets}

\begin{figure*}
   \sidecaption
   \includegraphics[width=12cm]{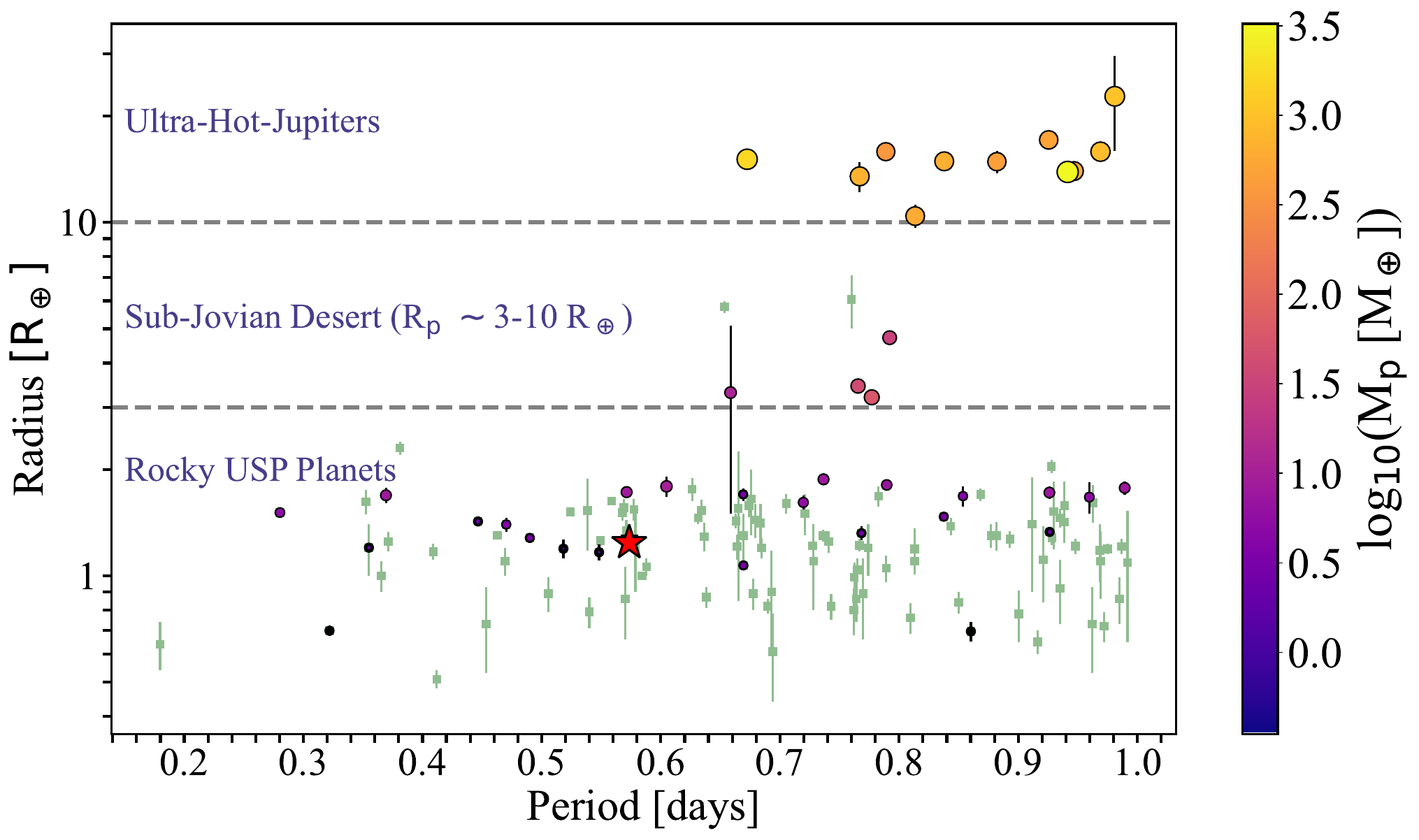}
   \caption{Orbital period versus radius diagram for ultra-short-period planets. Confirmed planets with only radius measurements are shown with green squares, and planets with mass and radius values are shown with circles. The color and size of the circles represent the mass of the planet. The position of Wolf~327b is marked by the red star. Data taken from NASA Exoplanet Archive (\url{https://exoplanetarchive.ipac.caltech.edu/}).}
   \label{Fig:USP_PeriodvsRadii}
\end{figure*}

Once we confirmed the planetary nature of Wolf~327b, we studied how it compares to the known population of USP planets. Figure \ref{Fig:USP_PeriodvsRadii} shows a period versus radius diagram of known exoplanets with short orbital periods ($P<1$ d), similar to the figures presented in the studies of \cite{Adams2021} and \cite{Morello2023}. The planets with radius and mass measurements are represented with circles, the size and color of the symbols being proportional to the mass of the planet; planets with only radius measurements are marked by the green squares. The figure highlights the range of sizes presented by USP planets: from Jupiter-sized planets, to a less populated sub-Jovian region, and with a more populated region inhabited by small ($R_\mathrm{p}<3 \; R_\oplus$) planets. The radius of Wolf~327b ($R_\mathrm{p} = 1.24 \pm 0.06 \; R_\oplus$) places it well below the sub-Jovian desert and in a region where the typical mass range goes up to a few tens of Earth masses, in agreement with our measured planetary mass. 

\begin{figure*}
    \sidecaption
    \includegraphics[width=12cm]{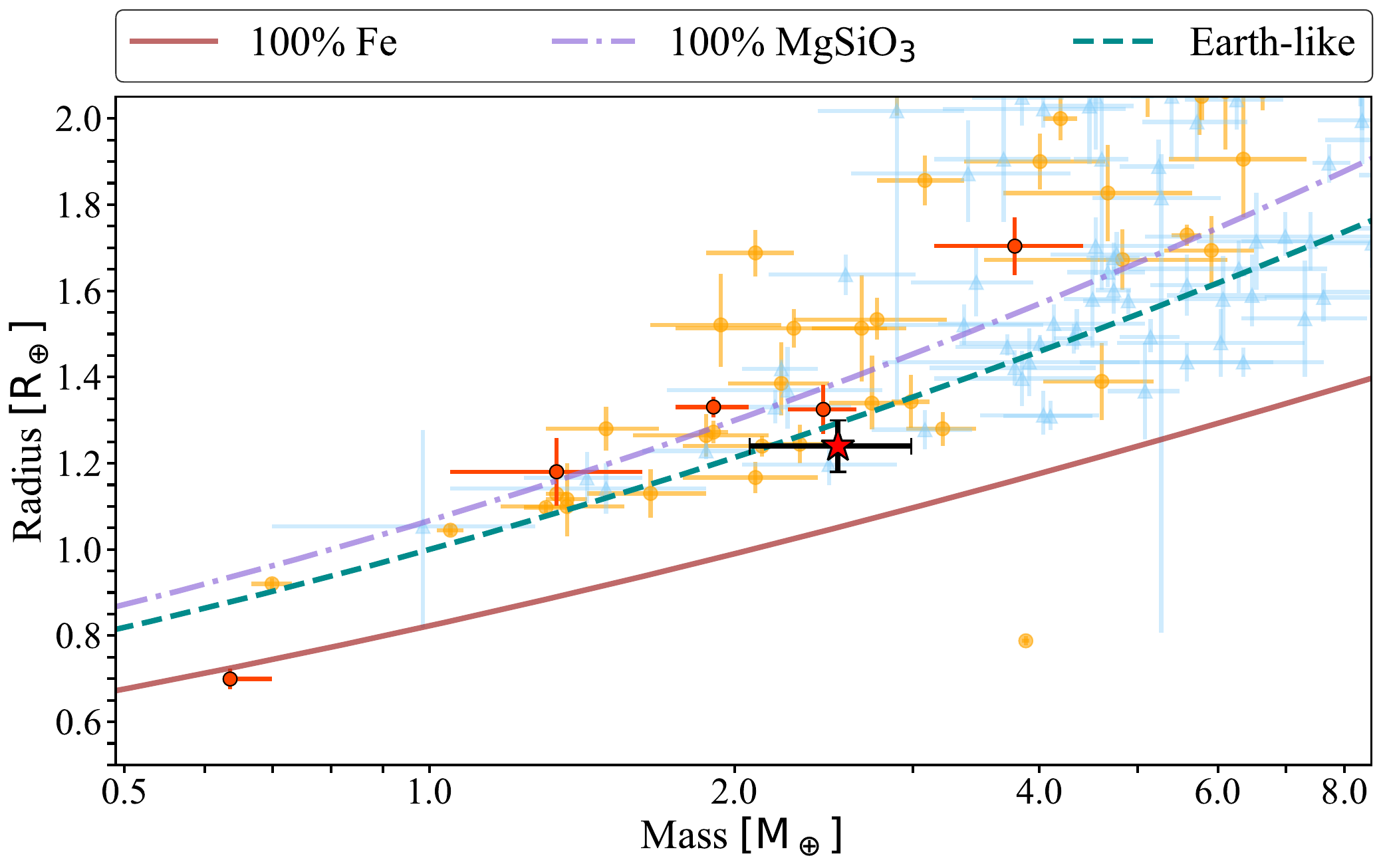}
    \caption{Mass--radius diagram for Wolf~327b (red star) and known transiting planets with mass determinations with a precision better than 30\% (parameters taken from the TEPcat database; \citealp{Southworth2011}). Planets orbiting M-type stars ($T_{\mathrm{eff}} \leq 4000$\,K) are marked with circles, orange indicate planets with periods $P > 1$ d, and red circles indicate USP planets ($P \leq 1$\,d) around M dwarfs. The lines in the mass--radius diagram represent the composition models of \cite{Zeng2016, Zeng2019} for planets with pure iron cores (100\% Fe, solid brown line), Earth-like rocky compositions (32.5\% Fe plus 67.5\% MgSiO$_3$, dashed green line) and pure rock (100\% MgSiO$_3$, dash-dotted purple line).}
    \label{Fig:MassRadius}
\end{figure*}

Figure \ref{Fig:MassRadius} shows the mass and radius values of known transiting exoplanets (data taken from TEPcat\footnote{\url{https://www.astro.keele.ac.uk/jkt/tepcat/}} catalog; \citealp{Southworth2011}). In the figure we also show the composition models of \cite{Zeng2016, Zeng2019}. As expected for small USP planets, Wolf~327b's position in this diagram indicates that this planet probably has a rocky composition, with little to no atmosphere, and with a higher density than that corresponding to an Earth-like composition for the planet's range of mass and radius. 

In terms of mass and radius, the USP planet with the most resemblance to Wolf~327b is K2-229b. This planet was discovered by \cite{Santerne2018}, who measured an orbital of period of $P=0.58$ d (13.9 h) and estimated a planetary radius of $R_\mathrm{p} = 1.17 \pm 0.07 \; R_\oplus$, a mass of $M_\mathrm{p} = 2.59 \pm 0.43 \; M_\oplus$ and a bulk density of $\rho_\mathrm{p} = 8.9 \pm 2.1$ g cm$^{-3}$. K2-229b is the inner planet of a planetary system, with a second planet in a 8.32 d orbit (K2-229c, $R_\mathrm{p} = 2.12 \pm 0.11 \; R_\oplus$, $M_\mathrm{p} < 21.3 \; M_\oplus$) and a possible third transiting candidate in a 31 d orbit (K2-229d, $R_\mathrm{p} = 2.65 \pm 0.24 \; R_\oplus$, $M_\mathrm{p} < 25.1 \; M_\oplus$). Incidentally, given the trend observed in the CARMENES RV data, Wolf~327b might also be part of a multiplanet system. Although K2-229b is similar in terms of size and mass, this planet has an equilibrium temperature of $T_{\rm eq} = 1960 \pm 40$ K because it is orbiting a star hotter than Wolf~327, a K0 dwarf ($T_{\rm eff} = 5185 \pm 32 $K). \cite{Santerne2018} modeled the planetary interior composition of K2-229b with only a metallic core and a silicate mantle, and found a core-mass fraction of 68\%, making K2-229b a planet with an enlarged core in comparison with its mantle, similar to Mercury. In the following section we explore the composition of Wolf~327b using theoretical planetary interior models.

\subsection{Planet interior modeling}
We utilized \texttt{ExoMDN}\footnote{\url{https://github.com/philippbaumeister/ExoMDN}} \citep{BaumeisterTosi2023} to model the interior of Wolf~327b, using as input our derived values for the planetary mass, radius, and equilibrium temperature while taking into account their uncertainties. \texttt{ExoMDN} is an exoplanet interior inference model that adopts a machine-learning approach. The program uses a set of mixture density networks (MDN) trained on 5.6 million synthetic planet models computed with the \texttt{TATOOINE} code \citep{Baumeister2020,MacKenzie2023}. The planets from the training set have masses below 25 $M_\oplus$, equilibrium temperatures in the range 100--1000\,K, and have compositions consisting of an iron core, a silicate mantle, a water and high-pressure ice layer, and an H/He atmosphere. Since our derived equilibrium temperature for Wolf~327b ($T_{\rm eq} = 996 \pm 23$\,K) places the planet at the edge of temperature range of \texttt{ExoMDN's} training dataset, we recommend that the results obtained by this model should be considered just as an approximation.

\begin{figure}
    \centering
    \includegraphics[width=\hsize]{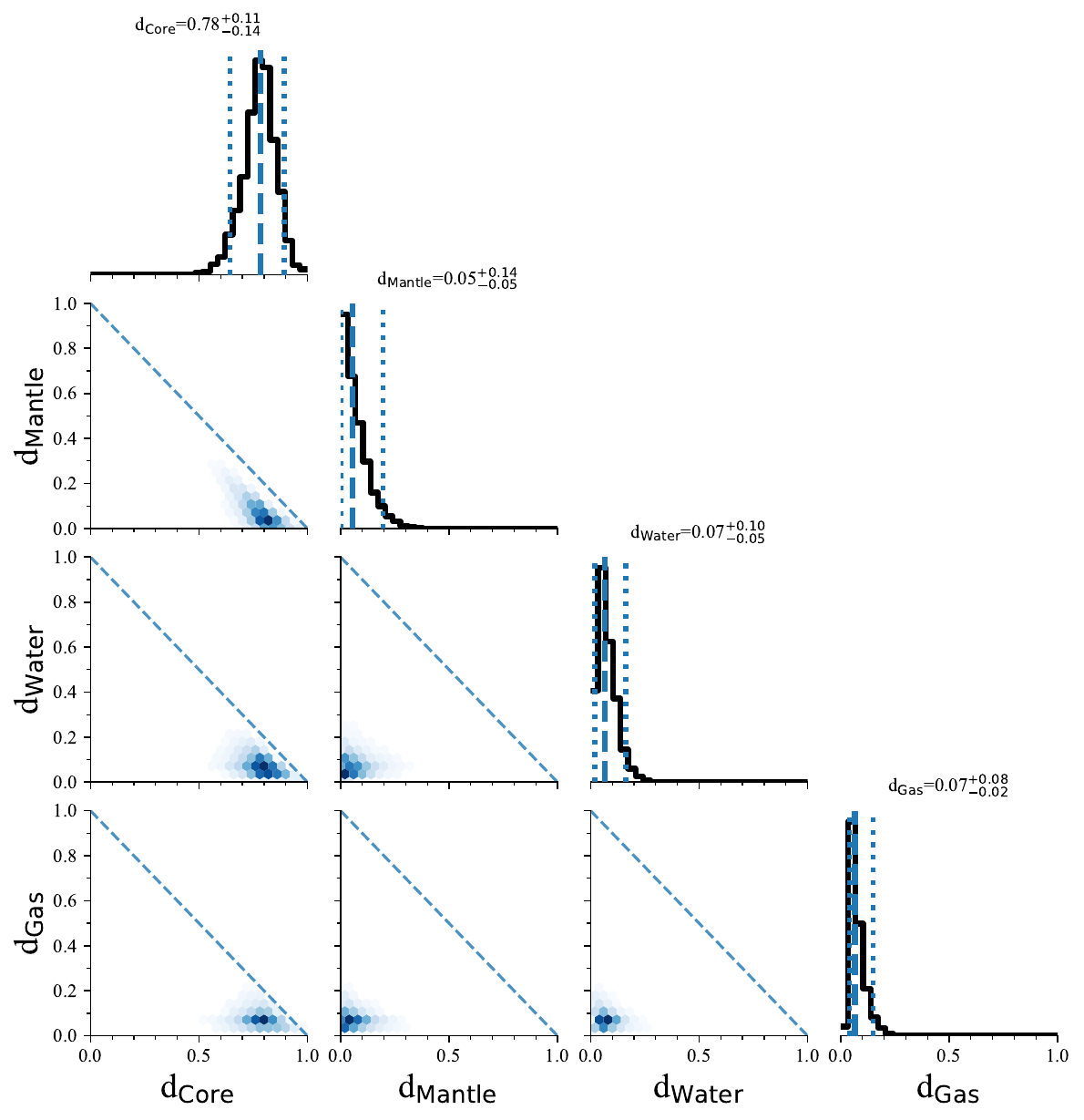}
    \caption{Predicted thickness of the interior layers of Wolf~327b computed with \texttt{ExoMDN}.}
    \label{Fig:Wolf327b_ExoMDN_RadiusFraction}
\end{figure}

\begin{figure}
    \centering
    \includegraphics[width=\hsize]{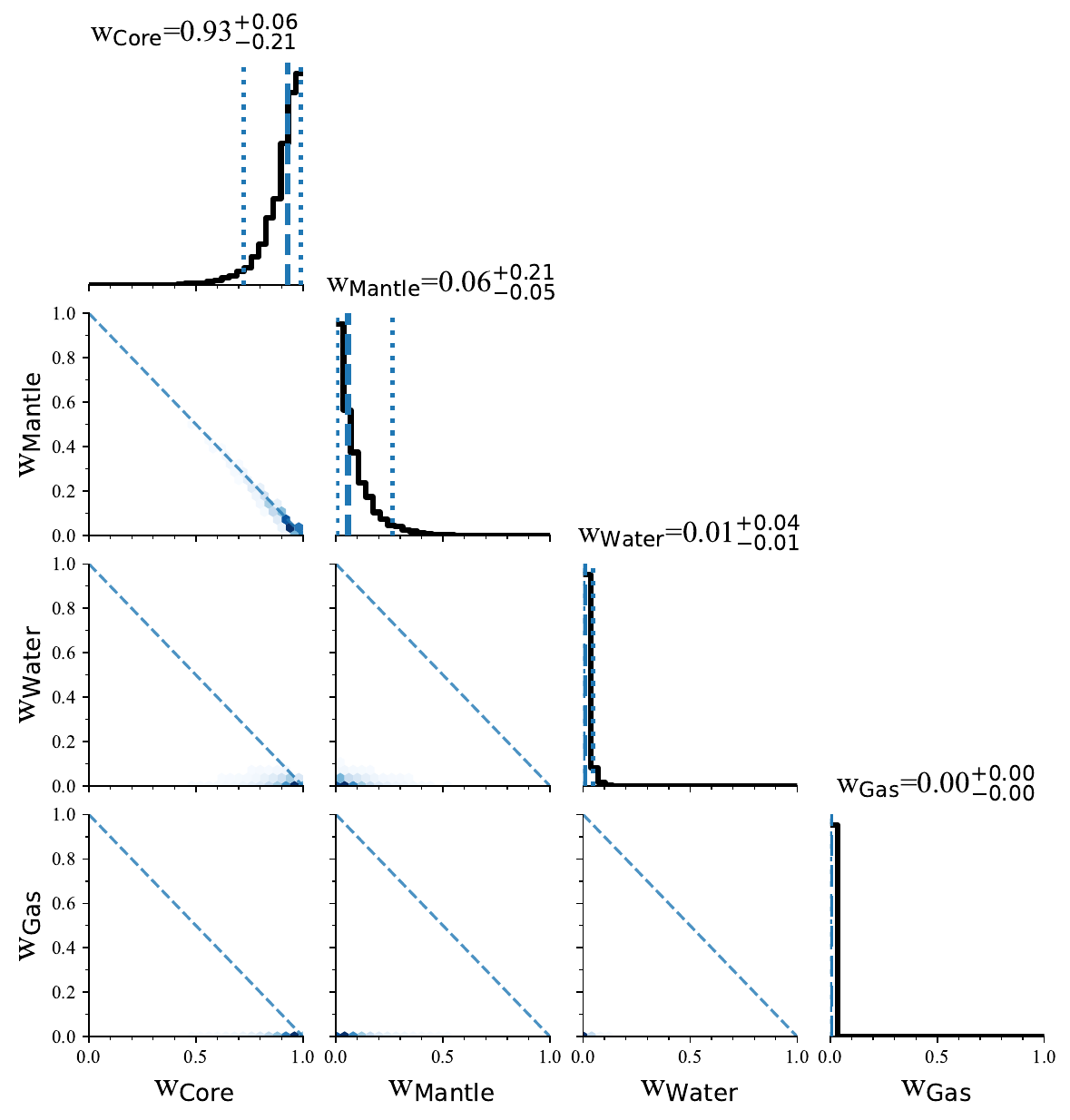}
    \caption{Predicted mass fraction of the interior layers of Wolf~327b computed with \texttt{ExoMDN}.}
    \label{Fig:Wolf327b_ExoMDN_MassFraction}
\end{figure}

Figures \ref{Fig:Wolf327b_ExoMDN_RadiusFraction} and \ref{Fig:Wolf327b_ExoMDN_MassFraction} present the thickness and mass fraction of the interior layers of Wolf~327b computed with \texttt{ExoMDN}. The interior model predicts that the planet is dominated in terms of layer size and mass fraction by a large planetary iron core; Wolf~327b's core could take up as much as 78\% of the planet's size and 93\% of its mass. For the other layers the model predicts a relatively small mantle, and a negligible fraction of water and gas layers. The predicted lack of an atmosphere is in agreement with what is expected for highly irradiated planets of relatively small sizes. Interestingly, \texttt{ExoMDN}'s prediction of a relatively large planetary core is in agreement with the simulations made for K2-229b in \cite{Santerne2018}, thus confirming the similarities between these two planets.

\subsection{Prospects for atmospheric characterization}
Studying the prerequisites for habitability and the orbital limits of the habitable zone requires the characterization of both temperate and hot planets~\citep[e.g.,][]{Turbet2019,Schlecker2023}. With $S_{\rm p} = 233.9 \pm 18.3 \; S_\oplus$, Wolf~327b is well within the latter regime~\citep{Kopparapu2013,Kopparapu2014}. If the planet accreted significant amounts of water or other volatiles during its formation, it probably went through a runaway greenhouse phase with an inflated atmosphere~\citep{Ingersoll1969,Kasting1988,Turbet2020,Mousis2020}, in which water is removed through photolysis and subsequent hydrodynamic escape of the hydrogen~\citep{Wordsworth2013,Wordsworth2014}. The expected past evolution of the planet's instellation~\citep{Baraffe2015} makes it likely that the planet never left this phase before being completely desiccated~\citep{Luger2015}. Our density estimate of Wolf~327b (see Fig.~\ref{Fig:MassRadius}) is consistent with a post-runaway greenhouse state without an extended atmosphere~\citep{Turbet2020}, which may have been further depleted through intense irradiation at 
the orbital distance of the planet~\citep{Owen2013}.

To evaluate the possibility of further detailed studies of this planet we computed the transmission spectroscopy metric (TSM) and emission spectroscopy metric (ESM) proposed by \cite{Kempton2018}. Both metrics are used to evaluate the feasibility of detecting atmospheric signals during a primary transit (TSM) or detecting the eclipse (ESM) with JWST. For Wolf~327b we find a TSM of $13$, \cite{Kempton2018} recommends a threshold of $\mathrm{TSM} > 10$ for planets with radii of $R_\mathrm{p} < 1.5 \; R_\oplus$ which marks this planet as a potential candidate for future transmission spectroscopy studies. However, owing to the high levels of irradiation suffered by Wolf~327b, it is possible that this planet has lost its original gas envelope. As for the possibility of eclipse detection, we find an ESM of $12.5$; the recommended ESM threshold for rocky planets is of $7.5$ which indicates that this planet is also a good candidate for eclipse spectroscopy with JWST. Both emission and transmission JWST observations would be a good way to test the hypothesis that the planet has little or no atmosphere.  

At the moment of submission, no JWST results on USP rocky planets had been reported. Early JWST transmission spectroscopy studies have focused on small planets with orbital periods longer than 1\,d and equilibrium temperatures below 1000\,K. These planets are: TRAPPIST-1b \citep[$P = 1.51$\,d, $R_\mathrm{p}= 1.11 \; R_\oplus$, and $T_{\rm eq} \approx 400$\,K;][]{Greene2023,Ih2023}, TRAPPIST-1c \citep[$P = 2.42$\,d, $R_\mathrm{p}= 1.09 \; R_\oplus$, and $T_{\rm eq} \approx 340$\,K;][]{Zieba2023,Lincowski2023}, LHS~475b \citep[$P = 2.03$\,d, $R_\mathrm{p}= 0.99 \; R_\oplus$, and $T_{\rm eq} \approx 590$\,K;][]{LustigYaeger2023}, and GJ~486b \citep[$P = 1.46$\,d, $R_\mathrm{p}= 1.30 \; R_\oplus$, and $T_{\rm eq} \approx 700$\,K;][]{Moran2023}. These planets are not considered to be USP planets and were expected to have some gaseous envelope. The observations for TRAPPIST-1b and 1c are consistent with a thin atmosphere or a bare rock surface, for LHS~475b JWST found a featureless transmission spectrum, and for GJ~486b the transmission spectrum can be explained by either the presence of water or by the effects of unocculted spots in the transmission spectrum. As discussed in Sect. \ref{Sec:Intro}, the USP rocky planet 55~Cnc~e is a prime candidate for JWST observations and has two approved Cycle 1 programs. \cite{Brandeker2021} will study whether the planet is in a 3:2 spin-orbit resonance to explain the secondary eclipse depth variability, and \cite{Hu2021} will try to detect any volatiles in the atmosphere of 5~Cnc~e using the spectral features of H$_2$O, CO, CO$_2$, and SiO. 

During the reviewing process of this article, \cite{Zhang2024} announced the first JWST results on a USP planet. The target of the study is GJ~367b \citep{Lam2021}, a USP sub-Earth with a radius and mass of $R_\mathrm{p}= 0.699 \pm 0.024 \; R_\oplus$ and $M_\mathrm{p}= 0.633 \pm 0.050 \; R_\oplus$ \citep{Goffo2023}, respectively, and an orbital period of 0.322\,d (7.7\,h). \cite{Zhang2024} measured the phase curve of GJ~367b in the mid-infrared wavelength range (5--12\,$\mu$m) using the Mid-Infrared Instrument \citep[MIRI;][]{Kendrew2015} on board of JWST. The team measured an eclipse depth of $79 \pm 4$\,ppm, with a lack of any detectable atmosphere, and featuring a dark surface (i.e., low albedo) within the MIRI bandpass. The measured emission spectrum is consistent with a blackbody spectrum. The measured low albedo value for GJ~367b is consistent with what is expected for lava worlds with molten surfaces \citep{Essack2020}.

\subsection{Planet-star interaction}
In this section we study the prospects for detecting radio emission from magnetic star-planet interaction between Wolf~327 and its newly discovered planet. The phenomenon behind this emission is the electron cyclotron maser (ECM) instability \citep{Melrose1982}, which can generate auroral radio emission in both the planet and its host star.
The characteristic frequency of this emission is given by the electron gyrofrequency, $\nu_G = 2.8 \, B$ (in MHz), with $B$ being the local magnetic field (in G). The resulting nonthermal emission has a coherent nature, with a high degree of circular polarization (close to 100\% in some instances), and can have a large bandwidth ($\Delta\,\nu \sim \nu_G/2$).

We note that any potential radio emission coming directly from the planet is most probably impossible to detect. Since the magnetic field of a planet is merely a few gauss, any resulting emission would fall well below the ionosphere cut-off. However, if the planet is in the sub-Alfv\'enic regime, where the speed of the plasma wind relative to the planet is smaller than the Alfv\'en speed, then energy and momentum can be transferred back to the star via Alfv\'en wings, and coherent electron-cyclotron emission can originate close to the star. Since the global magnetic fields of M-dwarf stars are typically larger than 100 G, sometimes reaching several kG, the electron gyrofrequency would be of a few hundred MHz, or even a few GHz, making the resulting auroral emission potentially detectable with current radio telescopes.

\begin{figure*}
    \centering
    \includegraphics[width=\hsize]{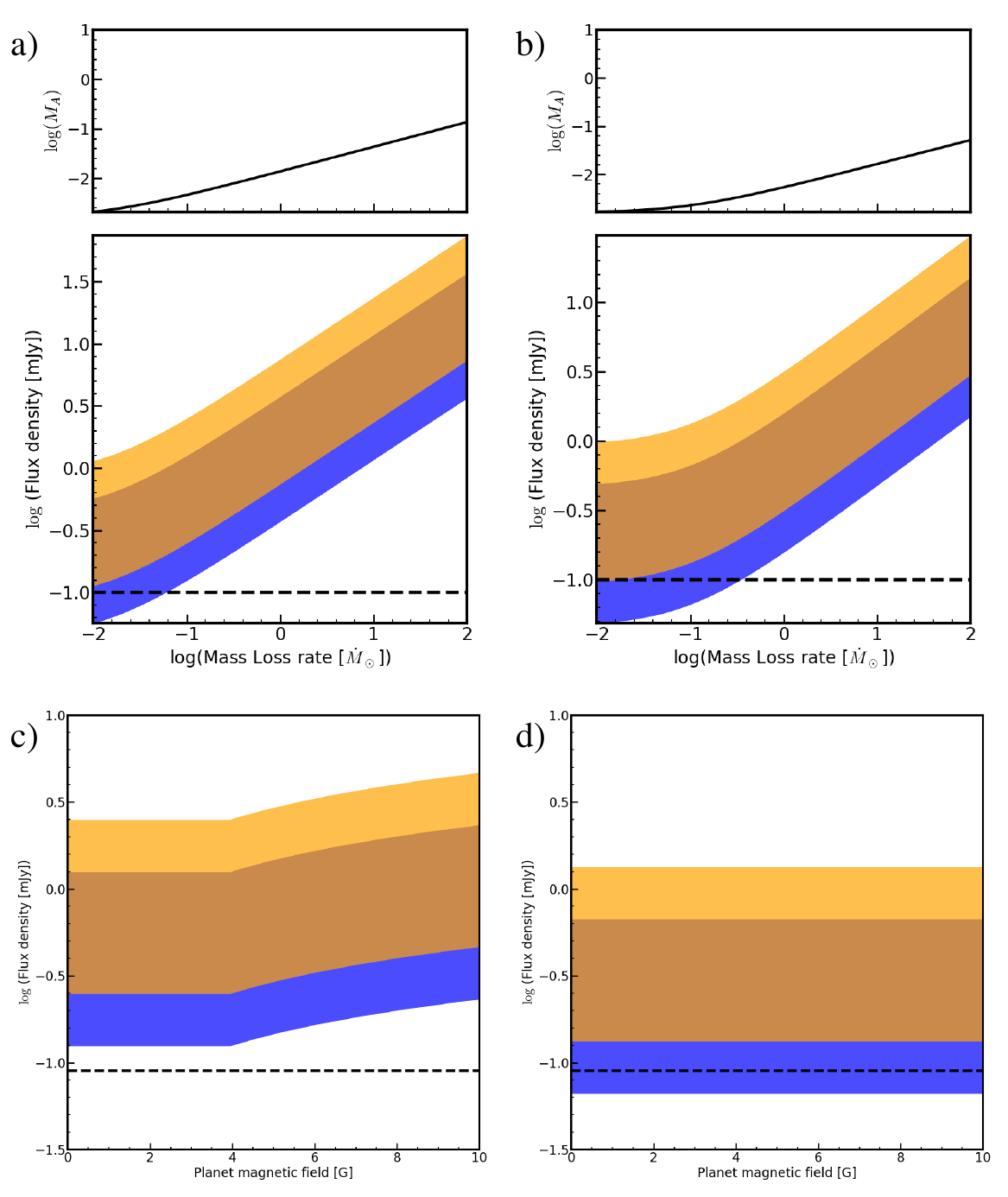}
    \caption{Flux density arising from star-planet interaction as a function of the stellar mass-loss rate (panels \textit{a} and \textit{b}) and the magnetic field of the planet (panels \textit{c} and \textit{d}). \textit{a)}: Expected flux density for a closed dipolar geometry. \textit{b)}: Expected flux density for an open Parker spiral geometry. \textit{c)}: Expected flux density for a closed dipolar geometry. \textit{d)}: Expected flux density for an open Parker spiral geometry. For  \textit{c)} and  \textit{d)} we assume $\dot{M}=0.1\dot{M}_\odot$. Note that the resulting radio emission is unaffected by the magnetic field of the planet, unless the latter is very large ($\sim$4\,G for closed dipole or even $> 10$\,G for a Parker spiral). The emission expected from Saur/Turnpenney's model is shown in orange, and the emission expected from Zarka/Lanza's model is shown in blue. The overlap between both models is shown in brown. The dashed black line represents the assumed detection threshold of $100\,\mu$Jy.}
    \label{Fig:StarPlanetInter}
\end{figure*}

We estimated the stellar magnetic field of Wolf~327 using the relations in \citet{Reiners2022}, which depend on the value of the Rossby number, $Ro$. From the two different $Ro$ estimates in \citet{Wright2018}, we get a magnetic field of the star, $B_\star \approx 292$\,G. To estimate the radio emission arising from star--planet interaction, we used the models in \citet{PerezTorres2021}, where we consider two different magnetic field geometries: a closed dipole and an open Parker spiral. We assumed that the stellar wind was isothermal, with $T = 3 \times 10^6$\,K, and that the solid angle covered by the ECM emission was 1.6 steradians. We considered two different emission models, Saur/Turnpenney's and Zarka/Lanza's, which are respectively shown as orange and blue in the plots. For each model we consider efficiency values ranging from 0.2 to 2\%. Since the planet is very close to the star, the interaction does indeed take place in the sub-Alfv\'enic regime (see top panels of Fig. \ref{Fig:StarPlanetInter}). 

We show in Fig.\ \ref{Fig:StarPlanetInter} (panels \textit{a} and \textit{b}) the radio flux density from sub-Alfv\'enic star--planet interaction as a function of the mass-loss rate of the star and find that the emission is strong and very similar in both a close dipolar and an open Parker spiral geometry of the magnetic field. Assuming a conservative threshold value of $100\,\mu\mathrm{Jy}$ for the rms obtained in a given observation, we should  be able to detect radio emission from star--planet interaction, regardless of the geometry, efficiency, model, or magnetic field of the planet.

\section{Conclusions}
\label{Sec:Conclusions}
We report the discovery of Wolf~327b, an ultra-short-period super-Earth orbiting around an M2.5~V dwarf with a period of 0.573\,d. The detection of the transit events was led by NASA's TESS mission, and ground-based observations confirmed that the transit events occur on the star Wolf~327. Follow-up RV measurements taken with the CARMENES spectrograph allowed us to establish the mass of the candidate and confirm its planetary nature.

From a coadded CARMENES mean stellar spectrum we determined stellar parameters for the host star. We estimate that Wolf~327 has an effective temperature of $T_{\rm eff} = 3542 \pm 70$\,K, an iron abundance of [Fe/H] $= -0.17 \pm 0.04$\,dex, a derived stellar mass of $M_\star = 0.405 \pm 0.019 \; M_\odot$, and a derived stellar radius of $R_\star = 0.406 \pm 0.015 \; R_\odot$. Using long-term ground-based photometric observations, we estimate a stellar rotation period of $P_{\rm rot} = 44.4 \pm 0.4$\,d. \textit{Gaia} DR3 measurements indicate the host star does not present any proper-motion companions, based on the star's Galactic velocity components it is likely that it does not belong to any young moving group association.

We analyzed photometric data taken by TESS (sectors 21 and 48), five ground-based individual full-transit observations, and 22 RV measurements taken with the CARMENES spectrograph. All the datasets were fitted simultaneously using an MCMC procedure that included the contribution of red noise sources for each time series modeled by Gaussian processes. We note that after subtracting the signal of the USP planet from the RV measurements, the residuals present a slope that might indicate the presence of an additional companion in the system. From our joint data fit we find that Wolf~327b is a super-Earth with a radius of $R_\mathrm{p} = 1.24 \pm 0.06 \; R_\oplus$ and mass of $M_\mathrm{p} = 2.53 \pm 0.46 \; M_\oplus$. These measurements imply that this planet has an iron-like bulk density of $\rho_\mathrm{p} = 7.24 \pm 1.66$\,g\,cm$^{-3}$. Owing to its short orbital period ($P=0.57347$\,d), the planet has an equilibrium temperature of $T_{\rm eq} = 996 \pm 22$\,K. Given its density and equilibrium temperature, it is likely that Wolf~327b possesses a rocky composition with little to no atmosphere. 

In terms of its mass and radius Wolf~327b, is very similar to the known transiting planet K2-229b. Previous studies of K2-229b indicate that this planet probably possesses a large iron core, with an internal structure similar to that of Mercury. Planet interior modeling applied to Wolf~327b arrives at a similar result: the internal structure of this planet is probably characterized by a substantial iron core, which dominates both in terms of layer size and mass fraction. Surrounding this core is a comparatively small mantle layer, accompanied by a negligible atmosphere (if any).

Overall, Wolf~327b is an interesting addition to the known USP planets around M dwarfs. This planet presents an interesting internal structure and favorable metrics for secondary transit detection with the JWST. Furthermore, since USP planets are usually found in multi-planetary systems, Wolf~327 also presents a good opportunity for additional RV follow-up to search for planets in wider orbits (as the data so far seem to indicate). 

\begin{acknowledgements}
    CARMENES is an instrument at the Centro Astron\'omico Hispano en Andaluc\'ia (CAHA) at Calar Alto (Almer\'{\i}a, Spain), operated jointly by the Junta de Andaluc\'ia and the Instituto de Astrof\'isica de Andaluc\'ia (CSIC). CARMENES was funded by the Max-Planck-Gesellschaft (MPG), the Consejo Superior de Investigaciones Cient\'{\i}ficas (CSIC), the Ministerio de Econom\'ia y Competitividad (MINECO) and the European Regional Development Fund (ERDF) through projects FICTS-2011-02, ICTS-2017-07-CAHA-4, and CAHA16-CE-3978, and the members of the CARMENES Consortium (Max-Planck-Institut f\"ur Astronomie, Instituto de Astrof\'{\i}sica de Andaluc\'{\i}a, Landessternwarte K\"onigstuhl, Institut de Ci\`encies de l'Espai, Institut f\"ur Astrophysik G\"ottingen, Universidad Complutense de Madrid, Th\"uringer Landessternwarte Tautenburg, Instituto de Astrof\'{\i}sica de Canarias, Hamburger Sternwarte, Centro de Astrobiolog\'{\i}a and Centro Astron\'omico Hispano-Alem\'an),  with additional contributions by the MINECO, the Deutsche Forschungsgemeinschaft (DFG) through the Major Research Instrumentation Programme and Research Unit FOR2544 ``Blue Planets around Red Stars'', the Klaus Tschira Stiftung, the states of Baden-W\"urttemberg and Niedersachsen, and by the Junta de Andaluc\'{\i}a.
    This paper includes data collected by the TESS mission, which are publicly available from the Mikulski Archive for Space Telescopes (MAST). Funding for the TESS mission is provided by NASA's Science Mission directorate. Resources supporting this work were provided by the NASA High-End Computing (HEC) Program through the NASA Advanced Supercomputing (NAS) Division at Ames Research Center for the production of the SPOC data products. We acknowledge the use of public TESS data from pipelines at the TESS Science Office and at the TESS Science Processing Operations Center. This research has made use of the Exoplanet Follow-up Observation Program website, which is operated by the California Institute of Technology, under contract with the National Aeronautics and Space Administration under the Exoplanet Exploration Program. 
    This work makes use of observations from the LCOGT network. Part of the LCOGT telescope time was granted by NOIRLab through the Mid-Scale Innovations Program (MSIP). MSIP is funded by NSF.
    The Joan Or\'{o} Telescope (TJO) of the Montsec Observatory (OdM) is owned by the Catalan Government and operated by the Institute for Space Studies of Catalonia (IEEC).
    This work is partly supported by JSPS KAKENHI Grant Number JP18H05439, JST CREST Grant Number JPMJCR1761.
    This article is based on observations made with the MuSCAT2 instrument, developed by ABC, at Telescopio Carlos S\'anchez operated on the island of Tenerife by the IAC in the Spanish Observatorio del Teide.
    This paper is based on observations made with the MuSCAT3 instrument, developed by the Astrobiology Center and under financial supports by JSPS KAKENHI (JP18H05439) and JST PRESTO (JPMJPR1775), at Faulkes Telescope North on Maui, HI, operated by the Las Cumbres Observatory.
    Some of the data presented herein were obtained at Keck Observatory, which is a private 501(c)3 nonprofit organization operated as a scientific partnership among the California Institute of Technology, the University of California, and the National Aeronautics and Space Administration. The Observatory was made possible by the generous financial support of the W. M. Keck Foundation. The authors wish to recognize and acknowledge the very significant cultural role and reverence that the summit of Maunakea has always had within the Native Hawaiian community. We are most fortunate to have the opportunity to conduct observations from this mountain.
    This research has made use of the Exoplanet Follow-up Observation Program (ExoFOP; DOI: 10.26134/ExoFOP5) website, which is operated by the California Institute of Technology, under contract with the National Aeronautics and Space Administration under the Exoplanet Exploration Program. 
    We acknowledge financial support from the Agencia Estatal de Investigaci\'on (AEI/10.13039/501100011033) of the Ministerio de Ciencia e Innovaci\'on and the ERDF ``A way of making Europe'' through projects 
    PID2022-137241NB-C4[1:4],	
    PID2021-125627OB-C3[1:2],	
    PID2019-109522GB-C5[1:4],	
    PID2019-107061GB-C61,       
    and RYC2021-031640-I,       
    and the Centre of Excellence ``Severo Ochoa'' and ``Mar\'ia de Maeztu'' awards to the Instituto de Astrof\'isica de Canarias (CEX2019-000920-S), Instituto de Astrof\'isica de Andaluc\'ia (CEX2021-001131-S) and Institut de Ci\`encies de l'Espai (CEX2020-001058-M).
    This work was also funded by the Generalitat de Catalunya/CERCA programme; 
    the M.\,V.~Lomonosov Moscow State University Program of Development;
    the Massachusetts Institute of Technology through the TESS mission via subaward s3449;
    the University of La Laguna under the EU Next Generation funds through the Margarita Salas Fellowship from the Ministerio de Universidades ref. UNI/551/2021-May 26;
    the DFG through the priority program SPP 1992 ``Exploring the Diversity of Extrasolar Planets'' (JE 701/5-1);
    the DFG through the Major Research Instrumentation Programme and Research Unit FOR2544 ``Blue Planets around Red Stars'';
    the Jet Propulsion Laboratory, California Institute of Technology, under contract 80NM0018D0004 with the National Aeronautics and Space Administration;
    the Francqui Foundation and the F.R.S.-FNRS through grant T.0109.20;
    and IPAC through a Visiting Research Fellowship. 
    The results reported herein benefited from collaborations and/or information exchange within NASA’s Nexus for Exoplanet System Science (NExSS) research coordination network sponsored by NASA’s Science Mission Directorate under Agreement No. 80NSSC21K0593 for the program ``Alien Earths''.

\end{acknowledgements}

%
%

\bibliographystyle{aa}
\bibliography{references}

\begin{appendix}
\section{Radial velocity data}
\label{Sec:RVdata}
Table \ref{Tab:CARMENES_RVs} present the RV measurements of Wolf~327 taken with the visible channel of CARMENES. Table presents \ref{Tab:CARMENES_Activ} the line activity indicators computed with \texttt{SERVAL}.

\begin{table*}[]
\caption{CARMENES VIS channel corrected radial velocities, instrumental drift, and nightly zero-point values computed by \texttt{SERVAL}. The RVs presented here were corrected by instrumental drifts and nightly zero-point offsets.}
\label{Tab:CARMENES_RVs}
\centering
\begin{tabular}{c c c c}
\hline 
\hline
BJD -- 2\,459\,000 & RV [m\,s$^{-1}$] & Instrumental drift [m\,s$^{-1}$] & Nightly zero-point [m\,s$^{-1}$]  \\
\hline
\noalign{\smallskip}

950.552461 & $4.615 \pm 1.618$ & $-0.737 \pm 0.442$ & $-1.108 \pm 1.015$ \\ 
952.431876 & $14.547 \pm 3.801$ & $2.820 \pm 0.465$ & $-1.198 \pm 1.677$ \\ 
952.657952 & $9.342 \pm 8.446$ & $-1.546 \pm 0.847$ & $-1.198 \pm 1.677$ \\ 
958.562133 & $4.469 \pm 1.829$ & $2.570 \pm 0.341$ & $-1.544 \pm 1.229$ \\ 
958.737953 & $5.320 \pm 2.080$ & $1.565 \pm 0.398$ & $-1.544 \pm 1.229$ \\ 
979.352322 & $2.662 \pm 2.464$ & $-9.406 \pm 0.455$ & $-1.813 \pm 0.501$ \\ 
979.393815 & $7.136 \pm 2.065$ & $-12.834 \pm 0.512$ & $-1.813 \pm 0.501$ \\ 
979.443206 & $6.017 \pm 2.354$ & $-16.900 \pm 0.590$ & $-1.813 \pm 0.501$ \\ 
979.492442 & $6.279 \pm 2.489$ & $-21.756 \pm 0.688$ & $-1.813 \pm 0.501$ \\ 
979.540070 & $6.173 \pm 2.073$ & $-26.460 \pm 0.542$ & $-1.813 \pm 0.501$ \\ 
979.586515 & $-0.143 \pm 1.839$ & $-30.637 \pm 0.454$ & $-1.813 \pm 0.501$ \\ 
979.633090 & $2.549 \pm 2.081$ & $-33.864 \pm 0.502$ & $-1.813 \pm 0.501$ \\ 
979.680690 & $-2.050 \pm 1.891$ & $-37.903 \pm 0.577$ & $-1.813 \pm 0.501$ \\ 
979.736000 & $-1.475 \pm 2.532$ & $-40.312 \pm 0.445$ & $-1.813 \pm 0.501$ \\ 
981.363716 & $-1.984 \pm 1.620$ & $-8.574 \pm 0.645$ & $-1.941 \pm 0.589$ \\ 
981.411369 & $-3.989 \pm 1.758$ & $-9.762 \pm 0.657$ & $-1.941 \pm 0.589$ \\ 
981.459496 & $-2.455 \pm 1.404$ & $-14.079 \pm 0.484$ & $-1.941 \pm 0.589$ \\ 
981.523068 & $-2.870 \pm 1.526$ & $-18.812 \pm 0.456$ & $-1.941 \pm 0.589$ \\ 
981.568274 & $-1.682 \pm 1.792$ & $-21.967 \pm 0.652$ & $-1.941 \pm 0.589$ \\ 
981.663415 & $1.686 \pm 1.636$ & $-29.524 \pm 0.448$ & $-1.941 \pm 0.589$ \\ 
981.697833 & $3.666 \pm 1.436$ & $-32.412 \pm 0.456$ & $-1.941 \pm 0.589$ \\ 
981.749403 & $2.931 \pm 1.866$ & $-35.338 \pm 0.476$ & $-1.941 \pm 0.589$ \\ 

\noalign{\smallskip}
\hline
\end{tabular}
\end{table*}

\begin{sidewaystable*}
\caption{CARMENES spectroscopic line activity indicators for Wolf~327 computed by \texttt{SERVAL}.}             
\label{Tab:CARMENES_Activ}      
\centering          
\begin{tabular}{l c c c c c c c c}     
\hline 
\hline
\noalign{\smallskip}
BJD -- 2\,459\,000 & CRX [m\,s$^{-1}$N$_\mathrm{p}^{-1}$] & dLW [m$^2$s$^{-2}$] & H$\alpha$ [m\,s$^{-1}$] & Na~{\sc i} D$_1$ [m\,s$^{-1}$] & Na~{\sc i} D$_2$ [m\,s$^{-1}$] & Ca~{\sc ii} IRTa [m\,s$^{-1}$] & Ca~{\sc ii} IRTb [m\,s$^{-1}$] & Ca~{\sc ii} IRTc [m\,s$^{-1}$]  \\
\hline
\noalign{\smallskip}

950.552461 & $17.6 \pm 12.3$ & $-3.2 \pm 1.3$ & $0.836 \pm 0.002$ & $0.213 \pm 0.005$ & $0.207 \pm 0.005$ & $0.590 \pm 0.002$ & $0.439 \pm 0.002$ & $0.412 \pm 0.002$ \\ 
952.431876 & $-34.6 \pm 35.4$ & $154.9 \pm 10.0$ & $0.792 \pm 0.005$ & $0.448 \pm 0.016$ & $0.388 \pm 0.016$ & $0.583 \pm 0.004$ & $0.427 \pm 0.005$ & $0.402 \pm 0.004$ \\ 
952.657952 & $33.5 \pm 88.5$ & $200.1 \pm 11.5$ & $0.755 \pm 0.013$ & $0.320 \pm 0.046$ & $0.280 \pm 0.048$ & $0.583 \pm 0.010$ & $0.448 \pm 0.012$ & $0.409 \pm 0.011$ \\ 
958.562133 & $-18.9 \pm 13.8$ & $-12.8 \pm 2.1$ & $0.828 \pm 0.003$ & $0.191 \pm 0.006$ & $0.177 \pm 0.006$ & $0.596 \pm 0.002$ & $0.440 \pm 0.002$ & $0.409 \pm 0.002$ \\ 
958.737953 & $16.7 \pm 17.6$ & $-5.2 \pm 2.1$ & $0.823 \pm 0.003$ & $0.342 \pm 0.009$ & $0.262 \pm 0.009$ & $0.580 \pm 0.003$ & $0.435 \pm 0.003$ & $0.408 \pm 0.003$ \\ 
979.352322 & $-35.4 \pm 26.0$ & $9.5 \pm 3.0$ & $0.869 \pm 0.004$ & $0.321 \pm 0.014$ & $0.274 \pm 0.015$ & $0.588 \pm 0.003$ & $0.449 \pm 0.003$ & $0.414 \pm 0.003$ \\ 
979.393815 & $-39.8 \pm 20.3$ & $-3.9 \pm 3.1$ & $0.845 \pm 0.004$ & $0.285 \pm 0.014$ & $0.202 \pm 0.015$ & $0.597 \pm 0.003$ & $0.448 \pm 0.003$ & $0.419 \pm 0.003$ \\ 
979.443206 & $26.3 \pm 23.7$ & $1.4 \pm 3.1$ & $0.841 \pm 0.005$ & $0.320 \pm 0.015$ & $0.282 \pm 0.015$ & $0.599 \pm 0.004$ & $0.446 \pm 0.004$ & $0.413 \pm 0.004$ \\ 
979.492442 & $52.3 \pm 23.6$ & $2.1 \pm 2.4$ & $0.844 \pm 0.004$ & $0.319 \pm 0.013$ & $0.267 \pm 0.013$ & $0.583 \pm 0.003$ & $0.448 \pm 0.003$ & $0.407 \pm 0.003$ \\ 
979.540070 & $10.9 \pm 20.8$ & $-9.5 \pm 2.1$ & $0.831 \pm 0.003$ & $0.262 \pm 0.010$ & $0.197 \pm 0.010$ & $0.591 \pm 0.003$ & $0.431 \pm 0.003$ & $0.408 \pm 0.003$ \\ 
979.586515 & $-24.2 \pm 18.1$ & $-15.2 \pm 2.6$ & $0.836 \pm 0.004$ & $0.201 \pm 0.010$ & $0.166 \pm 0.010$ & $0.598 \pm 0.003$ & $0.444 \pm 0.003$ & $0.413 \pm 0.003$ \\ 
979.633090 & $-0.6 \pm 21.2$ & $-15.7 \pm 2.2$ & $0.844 \pm 0.004$ & $0.247 \pm 0.012$ & $0.197 \pm 0.012$ & $0.595 \pm 0.003$ & $0.436 \pm 0.003$ & $0.416 \pm 0.003$ \\ 
979.680690 & $9.9 \pm 18.6$ & $3.0 \pm 2.7$ & $0.831 \pm 0.004$ & $0.372 \pm 0.013$ & $0.304 \pm 0.013$ & $0.592 \pm 0.004$ & $0.438 \pm 0.004$ & $0.413 \pm 0.003$ \\ 
979.736000 & $-19.2 \pm 26.8$ & $6.1 \pm 3.2$ & $0.838 \pm 0.004$ & $0.462 \pm 0.014$ & $0.323 \pm 0.014$ & $0.587 \pm 0.003$ & $0.450 \pm 0.004$ & $0.412 \pm 0.003$ \\ 
981.363716 & $0.9 \pm 15.1$ & $2.0 \pm 1.8$ & $0.829 \pm 0.003$ & $0.261 \pm 0.009$ & $0.223 \pm 0.009$ & $0.592 \pm 0.002$ & $0.436 \pm 0.002$ & $0.417 \pm 0.002$ \\ 
981.411369 & $-16.0 \pm 16.2$ & $1.4 \pm 1.3$ & $0.838 \pm 0.002$ & $0.215 \pm 0.005$ & $0.199 \pm 0.005$ & $0.593 \pm 0.002$ & $0.439 \pm 0.002$ & $0.411 \pm 0.002$ \\ 
981.459496 & $6.8 \pm 12.5$ & $-5.3 \pm 1.2$ & $0.821 \pm 0.002$ & $0.171 \pm 0.005$ & $0.170 \pm 0.005$ & $0.582 \pm 0.002$ & $0.436 \pm 0.002$ & $0.413 \pm 0.002$ \\ 
981.523068 & $3.9 \pm 14.2$ & $-8.8 \pm 1.4$ & $0.824 \pm 0.002$ & $0.189 \pm 0.005$ & $0.168 \pm 0.005$ & $0.584 \pm 0.002$ & $0.439 \pm 0.002$ & $0.414 \pm 0.002$ \\ 
981.568274 & $12.1 \pm 16.6$ & $-6.6 \pm 2.0$ & $0.816 \pm 0.003$ & $0.192 \pm 0.006$ & $0.179 \pm 0.006$ & $0.589 \pm 0.002$ & $0.439 \pm 0.002$ & $0.404 \pm 0.002$ \\ 
981.663415 & $-7.6 \pm 15.7$ & $-5.8 \pm 1.9$ & $0.819 \pm 0.003$ & $0.189 \pm 0.006$ & $0.171 \pm 0.006$ & $0.587 \pm 0.002$ & $0.440 \pm 0.002$ & $0.411 \pm 0.002$ \\ 
981.697833 & $-29.8 \pm 12.6$ & $-0.3 \pm 1.4$ & $0.820 \pm 0.002$ & $0.214 \pm 0.006$ & $0.190 \pm 0.006$ & $0.589 \pm 0.002$ & $0.439 \pm 0.002$ & $0.416 \pm 0.002$ \\ 
981.749403 & $30.3 \pm 18.4$ & $25.2 \pm 3.2$ & $0.822 \pm 0.003$ & $0.391 \pm 0.009$ & $0.301 \pm 0.009$ & $0.584 \pm 0.003$ & $0.434 \pm 0.003$ & $0.406 \pm 0.002$ \\ 

\noalign{\smallskip}
\hline                  
\end{tabular}
\end{sidewaystable*}

\section{Stellar rotation from TJO photometry}
\label{Sec:Appendix_TJO_photometry}
\begin{figure*}
   \centering
   \includegraphics[width=\hsize]{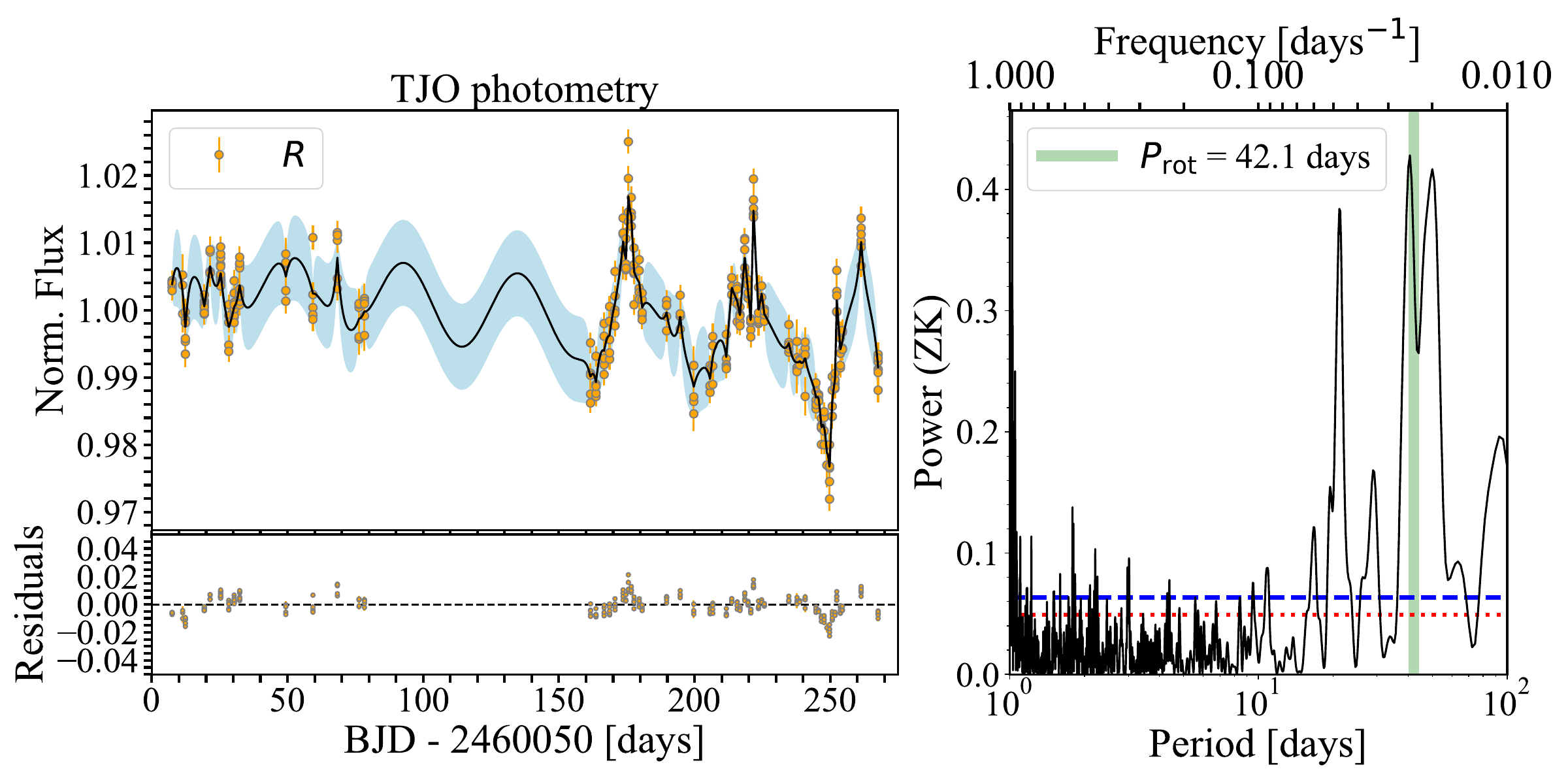}
   \caption{Long-term photometric observations of Wolf~327 taken with the Joan Or\'{o} telescope (TJO). \textit{Left panel:} normalized $R$-band flux (orange points) versus time. The best-fitted model is shown by the black line, the 1$\sigma$ uncertainty regions of the fit are shown in light blue. \textit{Right panel:} GLS periodogram \citep{Zechmeister2009} for TJO $R$-band photometry. The vertical green line shows the best-fitted rotation period ($P_{\mathrm{rot}}=42.1^{+3.5}_{-5.8} \; \mathrm{d}$). The horizontal lines represent the false alarm probability (FAP) levels of 10\% (red dotted line) and 1\% (blue dashed line).}
    \label{Fig:GroundBased_TJO_R}
\end{figure*}

In Fig. \ref{Fig:GroundBased_TJO_R} we show the ground-based photometric observations of Wolf~327 taken with the 0.8\,m Joan Or\'{o} telescope (TJO). On the left panel we present the normalized $R$-band flux versus time, in the right side panel the we show the GLS periodogram of the time-series. The periodogram presents a maximum peak at $P_{\mathrm{max}}=40.6 \; \mathrm{d}$. We attempted to fit the light curve using the quasi-periodic kernel described in Sec. \ref{Sec:StellarRotation}, however we found that by using this procedure the kernel was overfitting our data. We decided to model the light curve using a combination of a linear and a sinusoidal function:
 \begin{equation}
     L(t) = z_{\mathrm{pt}} + \alpha t + A\sin\left( \frac{2 \pi t}{P} + \phi \right)
  \label{Eq:TJO-lightcurve}
 ,\end{equation}
where $t$ is the epoch of the measurements, the free parameters $z_{\mathrm{pt}}$ and $\alpha$ are the zero point and slope of the linear function, respectively. For the sinusoidal function the free parameters are the amplitude ($A$), the rotation period ($P_{\mathrm{rot}}$), and a phase angle ($\phi$). To model the red noise present in the data, we used the GP implementation by \texttt{Celerite} \citep{ForemanMackey2017} and chose a simple exponential kernel of the form
\begin{equation}
    k_{ij \; \mathrm{exp}} = a^2 \exp^{-c (|t_i - t_j|)}
\label{Eq:TJO_GPKernel}
,\end{equation}
where $|t_i-t_j|$ is the time between epochs in the series, and the hyperparameters, $a$ and $c$, were allowed to be free.

The fitting procedure was the same as the one described in Sec. \ref{Sec:StellarRotation}. In Table \ref{Tab:Prot_parameters_tjo} we present the median and 1$\sigma$ uncertainties of the fitted parameters. With this procedure we find a rotation period of $P_{\mathrm{rot}} = 42.1^{+3.5}_{-5.8} \; d$, consistent with the rotation period found by fitting the ASAS-SN $V$-band data.

\begin{table}[t]
\centering
\caption{Prior functions and fitted parameters values for the TJO $R$-band photometry of Wolf~327.}
\label{Tab:Prot_parameters_tjo}
\renewcommand{\arraystretch}{1.3}
\begin{tabular}{l c c}
\hline 
\hline
\noalign{\smallskip}
Parameter & Prior & Value \\
\noalign{\smallskip}
\hline
\noalign{\smallskip}

$z_{\mathrm{pt}}$ & $\mathcal{U}(-2, 2)$ & $1.000^{+0.001}_{-0.001}$ \\
$\alpha \; (\times 10^{-6})$ [day$^{-1}$] & $\mathcal{U}(-50, 50)$ & $36.8^{+14.7}_{-14.6}$ \\
$A$ & $\mathcal{U}(10^{-6}, 10)$ & $0.0058^{+0.0017}_{-0.0023}$ \\
$P_{\mathrm{rot}}$ [d] & $\mathcal{U}(1.0, 100)$ & $42.1^{+3.5}_{-5.8}$ \\
$\phi$ [rad] & $\mathcal{U}(0, \pi)$ & $1.87^{+0.65}_{-1.09}$ \\
$\sigma_{\mathrm{phot\; jitter}}$ & $\mathcal{U}(0, 1)$ & $0.0015^{+0.0001}_{-0.0002}$ \\
$\log a$ & $\mathcal{U}(-13.8, 0.0)$ & $-10.1 \pm 0.3$ \\
$\log c$ [day$^{-1}$] & $\mathcal{U}(-4.6, 4.6)$ & $-0.5 \pm 0.5$ \\
\noalign{\smallskip}
\hline
\end{tabular}
\renewcommand{\arraystretch}{1}
\tablefoot{$\mathcal{U}$ represent a uniform prior function. The slope of the linear function ($\alpha$) was computed relative to the mid-time of the time series $T_{\mathrm{mid}} = 2460187$\,BJD.}
\end{table}

\section{Ground-based transit light curves}
\label{Sec:Appendix_GroundBasedLC}
\begin{figure*}
   \centering
   \includegraphics[width=\hsize]{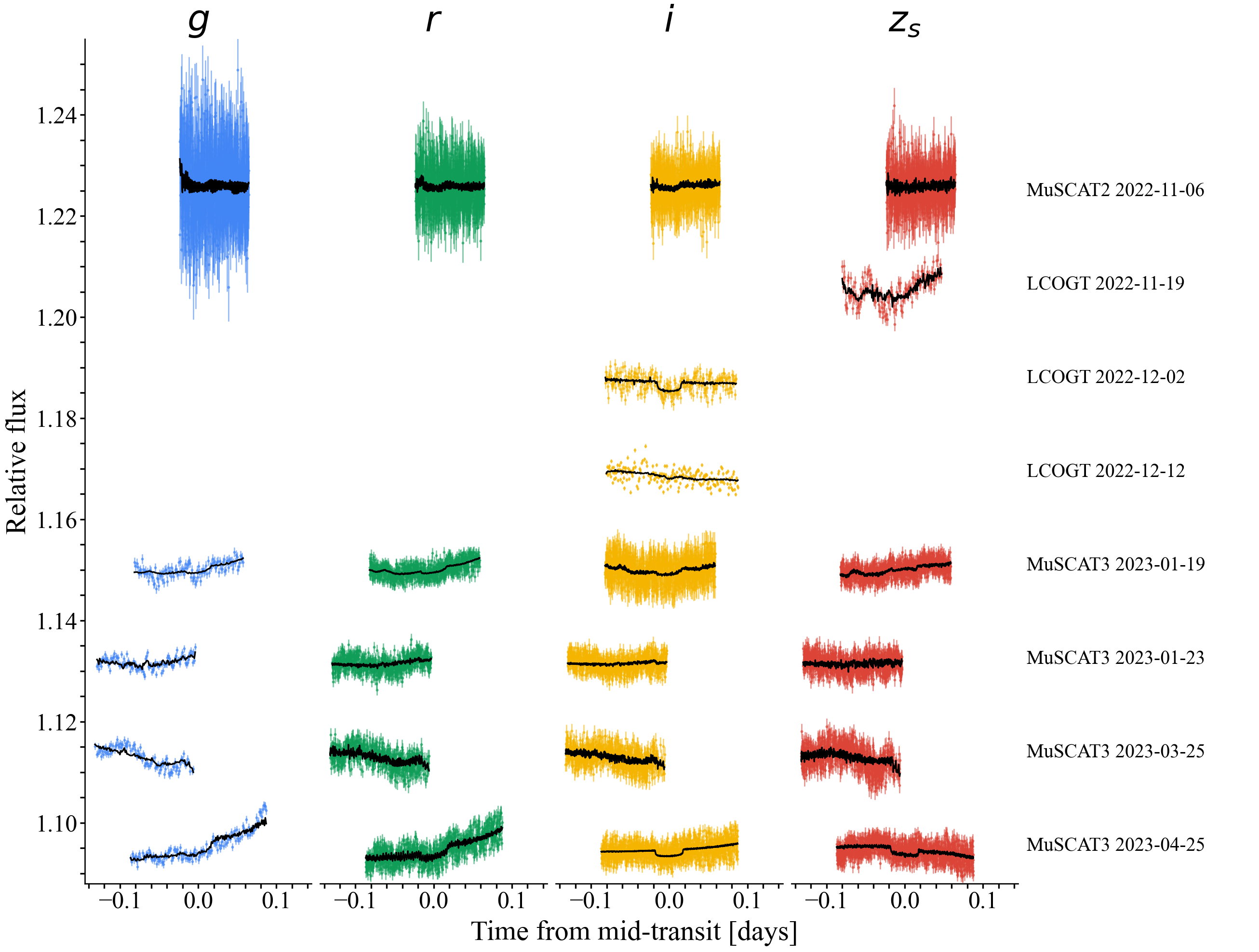}
   \caption{Ground-based transit light curves of Wolf~327b. The best fitted model (black line) includes systematic effects but no red noise component (i.e., without GPs).}
    \label{Fig:GroundBased_LightCurves_All}
\end{figure*}

In this section we describe our initial fitting process and selection for the ground-based light curves considered for the joint fit of the data. Figure \ref{Fig:GroundBased_LightCurves_All} shows all the ground-based transit observations available in ExoFOP and a time series taken with MuSCAT2, along with a transit model for each individual light curve (black line). The transit model was computed using normal priors in the central time of the transit and orbital period of the candidate, the values for these parameters were taken from the results of a test joint fit of the TESS and CARMENES datasets. To model some of the systematics present in the data we use a transit baseline taking into account the position of the star, the FWHM, and airmass during the observations:
\begin{equation}
    \mathcal{B} = f_0 + c_X X_p + c_Y Y_p + c_f F_w + c_a A_{ir}
    \label{Eq:Transit_ground_baseline}
,\end{equation}
where the free parameters are $f_0$ as a flux offset, the coefficients $c_X$ and $c_Y$ model the flux variations caused by changes in the star's position on the detector ($X_p, Y_p$), $c_f$ is the coefficient associated with the changes of the PSF's FWHM ($F_w$) during the observations, and $c_a$ is the free parameter used to model flux changes depending on the airmass ($A_{ir})$. For the individual transit fits we did not use GPs to model the red noise component present in each of the time series.

From Figure \ref{Eq:Transit_ground_baseline} we can see that the observations that cover a full transit event are from 2022 November 19, 2022 December 2 and 12, 2023 January 19, and 2023 May 25. These full transit observations were reanalyzed with {\tt AstroImageJ} using some transit parameter values as priors taken from a TESS and CARMENES joint fit. Then systematic instrumental effects were removed from the data, which were used for the final joint fit.


\section{Joint fit results}
\label{Sec:Appendix_JointFitResults}
\begin{figure}
    \centering
    \includegraphics[width=\hsize]{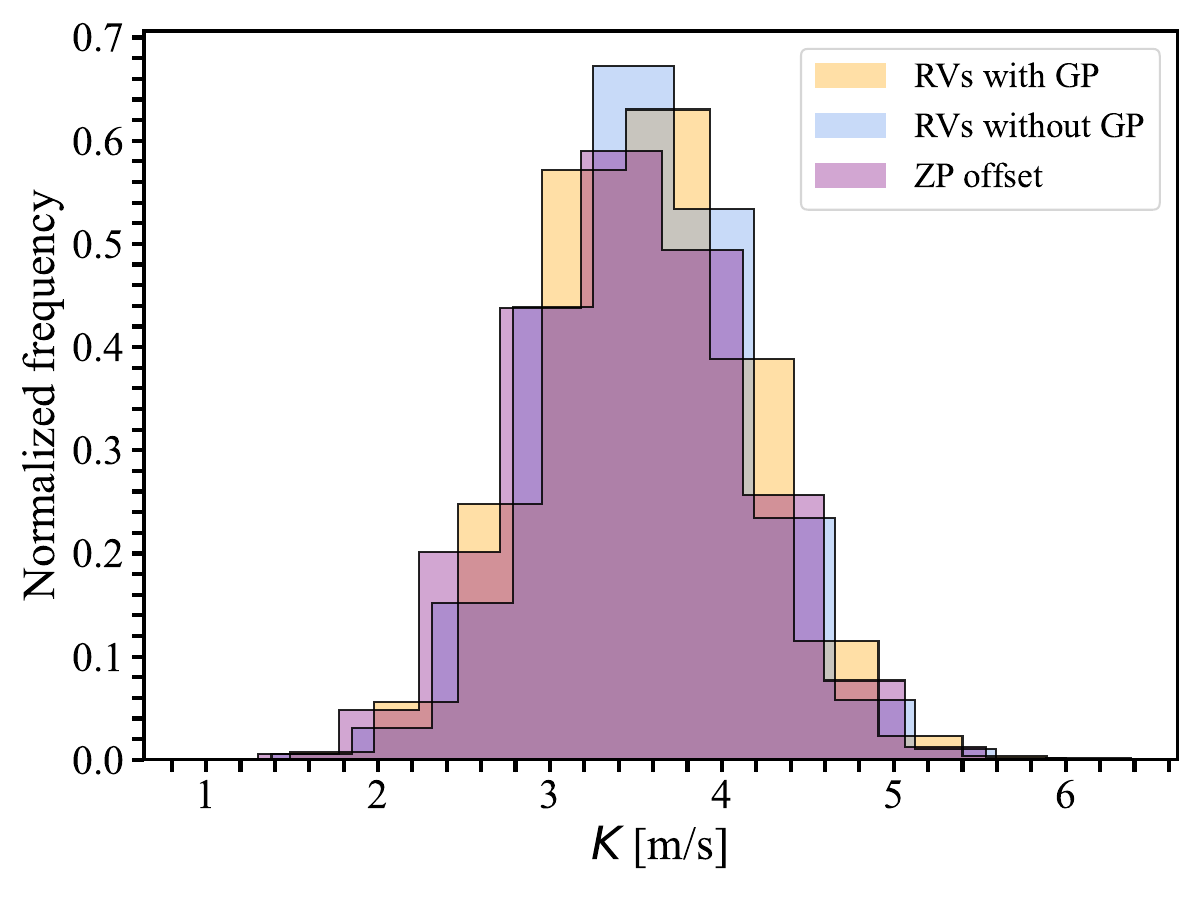}
    \caption{Posterior distribution of the radial velocity semi-amplitude ($K$) for a joint fit using only TESS and CARMENES RV observations. We show the results for a fit using GPs to model the red noise of both datasets (light yellow), GPs for TESS data and no GPs for CARMENES measurements (light blue), and a fit of TESS photometry including GPs and RVs without GPs but using a different zero-point (ZP) offset for each night (light purple).}
    \label{Fig:RVAmp_Test_GPs_vs_FloatingChunk}
\end{figure}

During our data fitting test phase we compared the results of a joint fit performed on TESS photometric data and CARMENES RV measurements but using different approaches to deal with the RVs. In one test we used GPs to model the red noise of both datasets (TESS and CARMENES), for a second model we used GPs for TESS and no GPs for CARMENES, and for the third model we used GPs only for TESS data and no GPs for CARMENES but assuming different zero-point offsets for each observed night. The zero-point offset method assumes that each night has a different radial velocity zero-point offset, meaning that we added five free $\gamma$ parameters corresponding to each night that the target was observed. Figure \ref{Fig:RVAmp_Test_GPs_vs_FloatingChunk} shows the posterior distribution of the radial velocity semi-amplitude $K$ obtained from all the models we tested, from the distributions we can see that either using or not using GPs, or the zero-point offset method, we get the same posterior distribution for $K$; meaning the that determination of the mass of the planet is unaffected by the use of GPs to model the systematics present in the RV measurements.

\begin{figure*}
   \centering
   \includegraphics[width=\hsize]{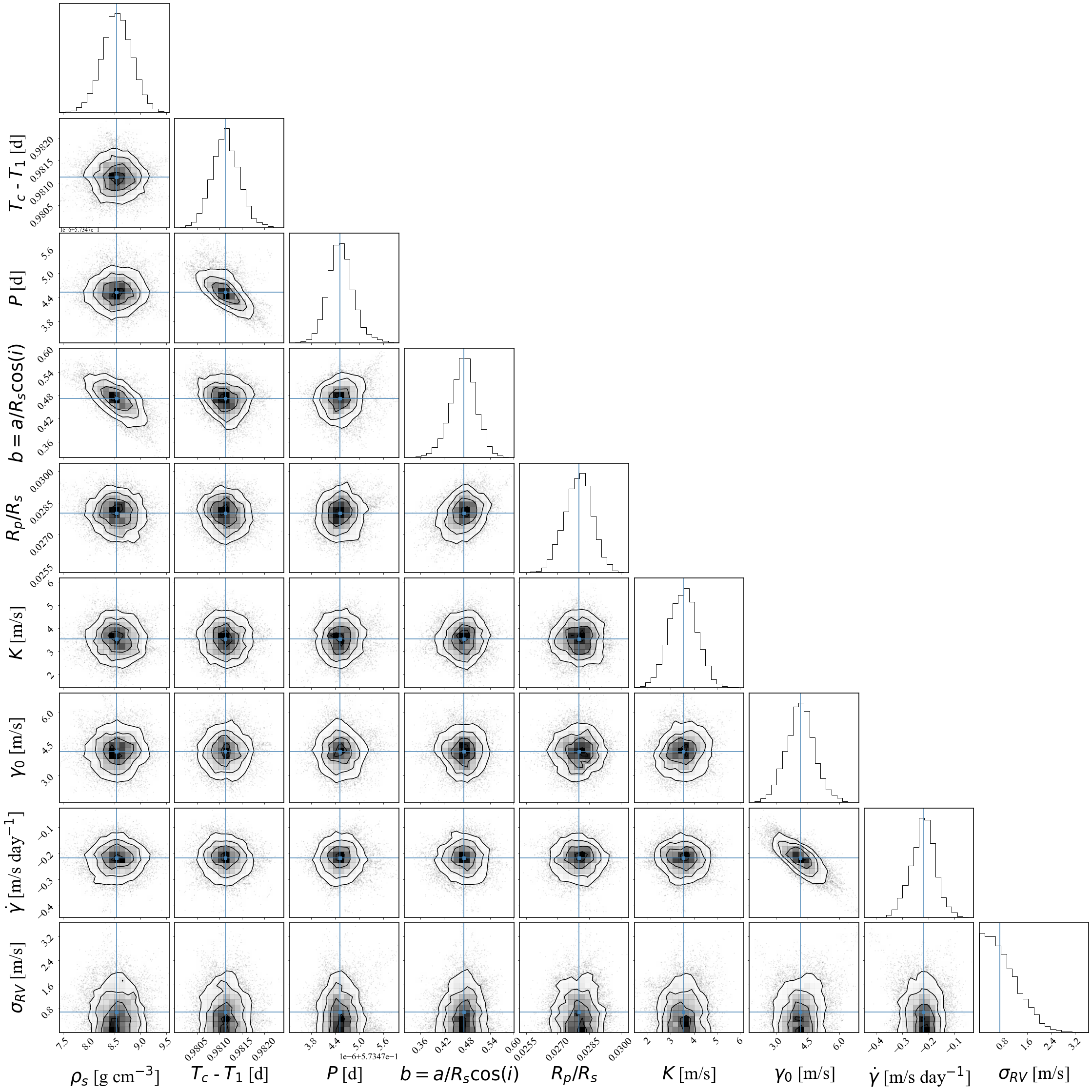}
   \caption{Correlation plot for the fitted transit and orbital parameter for Wolf~327b. The fitted orbital parameters of the eclipsing binary TIC~4918919 and the free parameters used to model the systematic effects were intentionally left out for easy viewing. The blue lines mark the median values of the distribution. For plotting purposes the distributions for the central time of the transit and orbital period were offset by T1 = 2459252 d.}
    \label{Fig:MCMC_CornerPlot}
\end{figure*}

Figure \ref{Fig:MCMC_CornerPlot} presents the posterior distributions for the fitted orbital parameters of Wolf~327b using all the data available (TESS, ground-based transit observations, and RV measurements from CARMENES). The fitted parameters for the eclipsing binary TIC~4918919 and Gaussian processes components were left out intentionally from the plot for easy viewing.

\end{appendix}

\end{document}